\documentclass{emulateapj}

\usepackage{natbib}

\shorttitle{Evolution of the Color--Density Relation}
\shortauthors{Cooper et al.}

\begin{document}


\title{The DEEP2 Galaxy Redshift Survey: Evolution of the
Color--Density Relation at $0.4 < \lowercase{z} < 1.35$}


\author{
Michael C.\ Cooper\altaffilmark{1},
Jeffrey A.\ Newman\altaffilmark{2,3},
Alison L.\ Coil\altaffilmark{4,3},
Darren J.\ Croton\altaffilmark{1},
Brian F.\ Gerke\altaffilmark{5},
Renbin Yan\altaffilmark{1},
Marc Davis\altaffilmark{1,5},
S.\ M.\ Faber\altaffilmark{6},
Puragra Guhathakurta\altaffilmark{6},
David C.\ Koo\altaffilmark{6},
Benjamin J.\ Weiner\altaffilmark{4,7},
Christopher N.\ A.\ Willmer\altaffilmark{4,8}
}

\altaffiltext{1}{Department of Astronomy, 
University of California at Berkeley, Mail Code
3411, Berkeley, CA 94720 USA; cooper@astro.berkeley.edu, 
darren@astro.berkeley.edu, renbin@astro.berkeley.edu, 
marc@astro.berkeley.edu}

\altaffiltext{2}{Lawrence Berkeley National Laboratory, 
1 Cyclotron Road Mail Stop 50-208,
Berkeley, CA 94720 USA; janewman@lbl.gov}

\altaffiltext{3}{Hubble Fellow}

\altaffiltext{4}{Steward Observatory, University of Arizona, 
933 N.\ Cherry Avenue, 
Tucson, AZ 85721 USA; acoil@as.arizona.edu, bjw@as.arizona.edu, cnaw@as.arizona.edu}

\altaffiltext{5}{Department of Physics, 
University of California at Berkeley, Mail Code
7300, Berkeley, CA 94720 USA; bgerke@astro.berkeley.edu}

\altaffiltext{6}{UCO/Lick Observatory, UC Santa Cruz, Santa Cruz, CA
95064 USA; faber@ucolick.org, 
raja@ucolick.org, koo@ucolick.org}


\altaffiltext{7}{Department of Astronomy, University of Maryland, 
College Park, MD 20742 USA}

\altaffiltext{8}{On leave from Observatorio Nacional, 
Rio de Janeiro, Brasil}

\begin{abstract}

  Using a sample of 19,464 galaxies drawn from the DEEP2 Galaxy Redshift
  Survey, we study the relationship between galaxy color and environment at
  $0.4 < z < 1.35$. We find that the fraction of galaxies on the red
  sequence depends strongly on local environment out to $z > 1$, being
  larger in regions of greater galaxy density. At all epochs probed, we
  also find a small population of red, morphologically early--type galaxies
  residing in regions of low measured overdensity. The observed
  correlations between the red fraction and local overdensity are highly
  significant, with the trend at $z > 1$ detected at a greater than
  $5\sigma$ level. Over the entire redshift regime studied, we find that
  the color--density relation evolves continuously, with red galaxies more
  strongly favoring overdense regions at low $z$ relative to their
  red--sequence counterparts at high redshift. At $z \gtrsim 1.3$, the red
  fraction only weakly correlates with overdensity, implying that any color
  dependence to the clustering of $\sim L^{*}$ galaxies at that epoch must
  be small. Our findings add weight to existing evidence that the build--up
  of galaxies on the red sequence has occurred preferentially in overdense
  environments (i.e., galaxy groups) at $z \lesssim 1.5$. Furthermore, we
  identify the epoch $(z \sim 2)$ at which typical $\sim L^{*}$ galaxies
  began quenching and moved onto the red sequence in significant number.
  The strength of the observed evolutionary trends at $0 < z < 1.35$
  suggests that the correlations observed locally, such as the
  morphology--density and color--density relations, are the result of
  environment--driven mechanisms (i.e., ``nurture'') and do not appear to
  have been imprinted (by ``nature'') upon the galaxy population during
  their epoch of formation.

\end{abstract}

\keywords{galaxies:high--redshift, galaxies:evolution, 
galaxies:statistics, galaxies: funamental parameters, 
large--scale structure of universe}

\section{Introduction}

The galaxy population both locally and out to $z \sim 1$ is found to be
effectively described as a combination of two distinct galaxy types: red,
early--type galaxies lacking much star formation and blue, late--type
galaxies with active star formation \citep[e.g.,][]{strateva01, baldry04,
  bell04, menanteau06}. The spatial distribution of this bimodal galaxy
population is frequently phrased today in terms of the so--called
morphology--density or color--density relation. As first quantified by
\citet{oemler74}, \citet{davis76}, and \citet{dressler80}, the
morphology--density relation holds that star--forming, disk--dominated
galaxies tend to reside in regions of lower galaxy density relative to
those of red, elliptical galaxies. 

Many physical mechanisms that could be responsible for this correlation
between galaxy morphology, star--formation history, and environment have
been proposed \citep[see][for a review of probable
mechanisms]{cooper06a}. Are the morphology-density and color--density
relations a result of environment--driven evolution, or were these trends
imprinted upon the galaxy population during their epoch of formation? Only
through comprehensive studies of galaxy properties and environments, both
locally and at high redshift, will we be able to understand the role of
local density in determining the star--formation histories and morphologies
of galaxies.

While the close relationship between galaxy type and density was primarily
uncovered via the study of nearby clusters, recent work using the 2--degree
Field Galaxy Redshift Survey \citep[2dFGRS,][]{colless01, colless03} and
the Sloan Digital Sky Survey \citep[SDSS,][]{york00} has established that
the connections between local environment and galaxy properties such as
morphology, color, and luminosity extend over the full range of densities,
from rich clusters to voids \citep[e.g.,][]{kauffmann04, balogh04a,
  blanton05, croton05, rojas05}. Furthermore, high--resolution imaging and
spectroscopic data in increasingly more distant clusters $({\rm to}\ z \sim
1)$ have shown that the trends observed locally persist to higher $z$, at
least in the highest density 
environments \citep[e.g.,][]{balogh97, treu03, poggianti06}.

Using a large sample of galaxies drawn from the DEEP2 Galaxy Redshift
Survey, \citet{cooper06a} extended the understanding of the color--density
relation at $z \sim 1$ across the full range of environments, from voids to
rich groups, showing that the correlation between galaxy color and mean
overdensity found locally is in place, at least in a global sense, when the
universe was half its present age. While the role of environment appears to
have been very critical at $z \sim 1$ and perhaps at earlier times,
quantitative measures of the evolution of environmental influences on the
galaxy population or of correlations such as the morphology--density and
color--density relation are limited.

Comparisons of local results with studies of high--redshift clusters have
pointed towards significant evolution in the relationship between galaxy
properties and local environment from $z \sim 1$ to $z \sim 0$
\citep[e.g.,][]{dressler97, couch98, smith05}. While these results indicate
an environment--driven evolution in the galaxy population (i.e., pointing
towards nurture versus nature as the origin of the galaxy bimodality), such
work has been limited to the vicinity of rich clusters, and thus we know
little about evolution in the relationship between galaxy type and density
across the full scope of galaxy environments. Furthermore, clusters include
only a relatively small fraction of the total galaxy population at any
epoch by number (and an even smaller fraction by volume). Thus, the
evolution of the color--density relation among the vast majority of the
galaxy population remains unprobed.

In large clusters, the physical mechanisms at work (e.g., galaxy
harassment, ram--pressure stripping, and global tidal interactions) go
beyond those acting in group--sized systems and the field. Results from the
first comprehensive study of galaxy environment over a broad range of
densities at high $z$ indicate that such cluster--specific physical
mechanisms cannot explain the global color--density relation as found at $z
\sim 1$ \citep{cooper06a}. Accordingly, in looking for evolution in the
relationship between the bimodal nature of galaxy properties and the local
galaxy environment, we must turn our attention to the entire dynamic range
of galaxy overdensities at high redshift.

In this vein, recent work employing a sample of low-- and high--redshift
field galaxies from the VIMOS VLT Deep Survey \citep[VVDS,][]{lefevre05}
has found a strong evolutionary trend in the color--density relation for
galaxies spanning the redshift range $0.25 < z < 1.5$, with the
color--magnitude diagrams for galaxies at $0.9 < z < 1.5$ showing no
significant dependence on environment \citep{cucciati06}. Also working at high
redshift, a study of the blue fraction (that is, the fraction of galaxies
that have blue color) in galaxy groups and in the field population by
\citet{gerke06b} finds instead that the field blue fraction significantly
differs from that of the group population out to $z \sim 1.3$.

In this paper, we use the large sample of high--$z$ galaxies obtained by the
DEEP2 survey to conduct a detailed study of the color--density relation at
$0.4 < z < 1.35$. In \S 2, we discuss the data sample employed along with
our measurements of galaxy environments and colors. Our main results
regarding the relationship between color and environment are presented in
\S 3. Finally, in \S 4 and \S 5, we discuss our findings alongside other
recent results and summarize our conclusions. Throughout this paper, we
assume a flat $\Lambda$CDM cosmology with $\Omega_m = 0.3$,
$\Omega_{\Lambda} = 0.7$, $w = -1$, and $h = 1$.

\section{The Data Sample}

\subsection{The DEEP2 Galaxy Redshift Survey}

The DEEP2 Galaxy Redshift Survey is a nearly completed $(>95\%)$ project
designed to study the galaxy population and large--scale structure at $z
\sim 1$ \citep{davis03, faber06}. To date, the survey has targeted $\sim \!
50,000$ galaxies in the redshift range $0.2 < z < 1.4$, down to a limiting
magnitude of $R_{\rm AB} = 24.1$. Currently, the survey covers $\sim \! 3$
square degrees of sky over four widely separated fields.

In this paper, we utilize a sample of 32,002 galaxies with accurate
redshifts \citep[quality ${\rm Q} = 3$ or ${\rm Q} = 4$ as defined
by][]{faber06} in the range $0.4 < z < 1.35$ and drawn from all four of the
DEEP2 survey fields. While at $z > 0.75$ the sample includes galaxies from
each of the four survey fields, spanning a total of $> \! 10$ pointings
\citep{faber06}, at $0.4 < z < 0.75$ our sample selection is limited to
galaxies in the Extended Groth Strip (EGS), where no color cut to
pre--select for $z > 0.7$ galaxies was used \citep{davis06}.

\subsection{Measurements of Rest--frame Colors, Luminosities, and Environments}

Rest--frame $(U-B)$ colors and absolute $B$--band magnitudes, $M_B$, are
calculated from CFHT $B,R,I$ photometry \citep{coil04b} using the
K--correction procedure described in \citet{willmer06}. All magnitudes
discussed within this paper are given in AB magnitudes \citep{oke83}. For
zero--point conversions between AB and Vega magnitudes, refer to Table 1 of
\citet{willmer06}.

For each galaxy in the data set, we compute the projected $3^{\rm
  rd}$--nearest--neighbor surface density $(\Sigma_3)$ about the galaxy,
where the surface density depends on the projected distance to the $3^{\rm
  rd}$--nearest--neighbor, $D_{p,3}$, as $\Sigma_3 = 3 / (\pi D_{p,3}^2)$. In
computing $\Sigma_3$, a velocity window of $\pm 1000\ {\rm km}/{\rm s}$ is
utilized to exclude foreground and background galaxies. In the tests of
\citet{cooper05}, this environment estimator proved to be the most robust
indicator of local galaxy density for the DEEP2 survey. To correct for the
redshift dependence of the sampling rate of the DEEP2 survey, each surface
density value is divided by the median $\Sigma_3$ of galaxies at that
redshift within a window of $\Delta z = 0.04$.; correcting the measured
surface densities in this manner converts the $\Sigma_3$ values into
measures of overdensity relative to the median density (given by the
notation $1 + \delta_3$ here) and effectively accounts for redshift
variations in the selection rate \citep{cooper05}.

\begin{figure}[h!]
\centering
\plotone{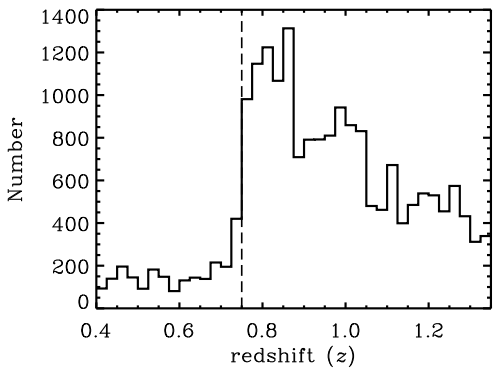}
\caption{The observed redshift distribution for the 19,464 galaxies in the
  four surveyed regions which are more than $1\ h^{-1}$ comoving Mpc away
  from a survey edge and within $0.4 < z < 1.35$. Galaxies at $z > 0.75$
  are selected from all four DEEP2 fields, while at $0.4 < z < 0.75$
  galaxies are drawn from the Extended Groth Strip only; in that field, no
  color--cut is used to pre--select high--redshift ($z > 0.7$) sources. In the
  figure, a dashed line is employed to illustrate this transition in the
  sample at $z = 0.75$. The redshift histogram is plotted using a bin size
  of $\Delta z = 0.025$.}
\label{zdist}
\end{figure}

Finally, to minimize the effects of edges and holes in the survey geometry,
we exclude all galaxies within $1\ h^{-1}$ comoving Mpc of a survey
boundary, reducing our sample from 32,002 to 19,464 galaxies. The redshift
distribution of this subset is plotted in Figure \ref{zdist}. For complete
details regarding the computation of the local environment measures, we
direct the reader to \citet{cooper06a}.

Within the DEEP2 sample, the bimodality of galaxy colors in rest--frame
$U-B$ color is clearly visible; out to $z \sim 1.4$, the galaxy
color--magnitude diagram exhibits a clear division into a relatively tight
red sequence and a more diffuse ``blue cloud'' of galaxies
\citep{willmer06}. For this study, we compute the fraction of galaxies on
the red sequence using the following color division, as defined by
\citet{willmer06} and illustrated as the dotted lines in Figure
\ref{magcut} and the dashed lines in the lower panels of Figure
\ref{redfrac}:
\begin{equation}
U - B = -0.032 (M_B - 21.62) + 0.454 - 0.25 + 0.831.
\label{willmer_cut}
\end{equation}
The red fraction $(f_{\rm R})$ within a given redshift and environment
range is given by the number of galaxies redward of this relation in $U-B$
color divided by the total number of galaxies within the same bin of
redshift and environment. 

As discussed by \citet{willmer06}, this division between the red sequence
and blue cloud is derived from the work of \citet{vandokkum00}, in which
the color--magnitude relation for early--type (i.e., solely
morphologically--selected) galaxies is measured. The results of this work
and similar studies of clusters at $z < 1$ show that ellipticals form a red
sequence in color--magnitude space with little scatter
\citep[e.g.,][]{ellis97, stanford98}. Thus, dividing the sample according
to the color--cut presented in Equation \ref{willmer_cut} effectively
selects the red, early--type portion of the bimodal galaxy
distribution. Details regarding the possible evolution of this color
division with redshift are highlighted by \citet{willmer06} and examined in
\S 3.1.1 of this work.

\subsection{Sample Selection}

Because the DEEP2 survey extends over a broad redshift range, selecting
galaxies according to a fixed apparent magnitude limit introduces
differences in selection with redshift that depend in a nonnegligible way
on galaxy color and luminosity. To study how these selection effects
potentially influence our results, we identify several subsamples within
the full catalog (Sample A) of 19,464 DEEP2 galaxies spread across the
redshift range $0.4 < z < 1.35$ (cf.\ Fig.\ 1). The simplest selection
method is to produce a volume--limited subsample according to a strict cut
in absolute magnitude. We create such a sample (Sample B) by restricting to
$0.4 < z < 1.1$ and requiring $M_B \le -21$, the absolute magnitude to
which DEEP2 is complete along both the red sequence and the blue cloud at
$z = 1.1$ (cf.\ top of Figure \ref{sample_select}). 

Producing a volume--limited sample with a fixed absolute--magnitude limit at
all colors severely restricts either the redshift range probed or the
number of galaxies selected at each redshift. Here, we adopt a limiting
redshift of $z = 1.1$ in order to maintain enough galaxies (2,784 in Sample
B) with which to accurately compute the red fraction over the full range of
overdensities. Given the large redshift range probed in our analysis,
however, the number of galaxies with $M_B \le -21$ at low redshift $(0.4 <
z < 0.75)$ is fairly small, and as such results at low $z$ using Sample B
are quite noisy.

As discussed by \citet{gerke06b}, in studying the evolution of galaxy
properties it is also possible to produce volume--limited catalogs with a
color--dependent, absolute--magnitude cut by defining a region of rest-frame
color--magnitude space that is uniformly sampled by the survey at all
redshifts of interest. For the DEEP2 survey, such a selection cut is
illustrated in the bottom of Figure \ref{sample_select} of this paper and
Figure 2 of \citet{gerke06b} and given by
\begin{equation}
M_{\rm{cut}}(z,U-B) = 
\begin{array}{l l}
Q(z - z_{\rm lim}) + \\
{\rm min}\left\{[a(U-B) + b],\ [c(U-B) + d]\right\}, 
\end{array}
\label{magcut}
\end{equation}
where $z_{\rm{lim}}$ is the limiting redshift beyond which the selected
sample becomes incomplete, $a$, $b$, $c$ and $d$ are constants that are
determined by the limit of the color--magnitude distribution of the sample
with redshift $z < z_{\rm{lim}}$, and $Q$ is a constant that allows for
linear redshift evolution of the typical galaxy absolute magnitude
$M_{B}^{*}$. For the parameter $Q$, we adopt the \citet{faber05} value of
$Q = -1.37$, determined from a study of the $B$--band galaxy luminosity
function in the COMBO--17 \citep{wolf01}, DEEP1 \citep{vogt05}, and DEEP2
surveys.

\begin{figure}[h!]
\centering
\plotone{f2.eps}
\caption{(\emph{Top}) The rest--frame color--magnitude diagram for galaxies
  in the redshift range $1.075 < z < 1.125$, selected from the full galaxy
  catalog (Sample A). The dotted vertical line defines the
  color--independent completeness limit of the DEEP2 survey at this redshift
  --- the absolute--magnitude limit used for Samples B \& D. To this limit,
  the samples are complete for galaxies of all color at $z < 1.1$
  (\emph{Bottom}) The rest--frame color--magnitude diagram for galaxies in
  the redshift range $1.275 < z < 1.325$, drawn from the full galaxy
  catalog (Sample A). The dotted line defines the completeness limit of the
  DEEP2 survey as a function of rest--frame color at redshift $z = 1.3$, as
  given by Equation \ref{magcut}. Due to the $R$--band magnitude limit of
  the survey, the fraction of red galaxies in the overall DEEP2 sample
  decreases significantly at high redshift, but by staying within this
  limit we can construct samples that are free of this effect. The dashed
  horizontal line shown in both plots illustrates the division between the
  red sequence and the blue cloud used throughout this paper, following the
  relation of \citet{willmer06} (cf.\ Equation \ref{willmer_cut}).  }
\label{sample_select}
\end{figure}

By including this linear $M_{B}^{*}$ evolution in our selection cut, we are
selecting a similar population of galaxies with respect to $M_{B}^{*}$ at
all redshifts. Adopting this approach with a limiting redshift of $z_{\rm
  lim} = 1.3$, we define a sample of 11,192 galaxies (Sample C) over the
redshift range $0.4 < z < 1.3$ that is volume--limited relative to
$M_{B}^{*}$ and selected according to this color--dependent cut in
$M_{B}$. The values of the constants $a$, $b$, $c$, and $d$ which define
the color--dependent selection are $-2.06$, $-18.9$, $-2.63$, and $-18.5$,
respectively. For complete details of the selection method, we refer the
reader to \citet{gerke06b}.

As a final sample, we again select galaxies assuming linear evolution in
$M_{B}^{*}$, as for sample C. However, in determining the limiting absolute
magnitude as a function of redshift, we do not apply a color--dependent
selection as utilized for Sample C. Instead, we restrict galaxies of all
color to the selection limit,
\begin{equation}
\begin{array}{c}
M_{B}(z) \le M_{\rm lim} - Q (z_{\rm lim} - z)\\ 
{\rm or} \\
M_{B}(z) - M_{*}(z) \le M_{\rm lim} - M_{*}(z_{\rm lim}), 
\end{array}
\label{magcut2}
\end{equation}
where the limiting redshift is chosen to be $z_{\rm lim} = 1.1$ and $M_{\rm
  lim} = -21$ defines the absolute magnitude at $z = 1.1$ to which DEEP2 is
complete along both the red sequence and the blue cloud (cf.\ the top panel
of Fig.\ \ref{sample_select}). Due to the more severe, color--independent
cut in absolute magnitude, Sample D totals only 2,150 galaxies, spanning
the redshift range $0.4 < z < 1.1$. In contrast, while Sample B is
similarly selected based just on absolute magnitude (that is, independent
of galaxy color), the magnitude cut in Sample D evolves with redshift,
staying fixed relative to $M_B^{*}$. Thus, Sample B selects galaxies down
to the same absolute magnitude at all redshifts, and Sample D samples
galaxies to the same depth in the luminosity function at each $z$. A brief
summary of the four galaxy samples utilized in this paper is provided in
Table \ref{sample_descript_tab}.

\begin{deluxetable*}{l l l l}
\tablewidth{0pt}
\tablecolumns{5}
\tablecaption{\label{var_tab} Descriptions of Galaxy Samples}
\tablehead{ Sample & $N_{\rm galaxies}$ & $z$ range & Brief Description }
\startdata
Sample A & 19,464 & $0.4 < z < 1.35$ & 
\parbox[l]{2.75in}{all galaxies after boundary cut} \\
Sample B & 2,784 & $0.4 < z < 1.1$ & 
\parbox[l]{2.75in}{color--independent, volume--limited $(M_B < -21)$ cut} \\
Sample C & 11,192 & $0.4 < z < 1.3$ & 
\parbox[l]{2.75in}{color--dependent limit, with limit held 
constant relative to $M_{B}^{*}(z)$} \\
Sample D & 2,150 & $0.4 < z < 1.1$ & 
\parbox[l]{2.75in}{color--independent limit, with limit held
constant relative to $M_{B}^{*}(z)$ and set as $M_{B} = -21$ at $z = 1.1$}\\ 
\enddata
\tablecomments{We list each galaxy sample employed in the analysis,
  detailing the selection cut used to define the sample as well as the
  number of galaxies $(N_{\rm galaxies})$ included and the redshift range
  covered by each sample.}
\label{sample_descript_tab}
\end{deluxetable*}

\subsection{Mock DEEP2 Survey Catalogs}

In order to test for possible systematic effects we employ a set of 12 mock
galaxy catalogs based on those of \citet{yan04}. These catalogs are derived
from $N$--body simulations by populating dark matter halos with galaxies
according to a halo occupation distribution (HOD) function
\citep{peacock00, seljak00}, which describes the probability distribution
of the number of galaxies in a halo as a function of the host halo
mass. The luminosities of galaxies are then assigned according to the
conditional luminosity function (CLF) formalism introduced by
\citet{yang03}, which allows the galaxy luminosity function to be
mass--dependent as well. Parameters for the HOD and the CLF are chosen to
match the 2dFGRS luminosity function \citep{madgwick02} and two--point
correlation function \citep{madgwick03}. By assuming that the manner in
which dark matter halos are populated with galaxies does not evolve from $z
\sim 1$ to $z \sim 0$ \citep{yan03} save via an overall evolution in
$M_B^{*}$ (corresponding to $Q=-1$ here), mock catalogs can then be built
using dark--matter--only simulation outputs at varying redshifts. The
resulting simulated galaxy catalogs from \citet{yan04} are in excellent
agreement with the lower redshift $(0.7 < z < 0.9)$ DEEP2 correlation
function \citep{coil04b} and the COMBO--17 luminosity function
\citep{wolf03b}.

It is critical for this paper that we characterize any systematic effects
which alter the relationship between galaxy color and observed overdensity
as a function of redshift. Therefore, we employ a modified version of the
\citet{yan04} mock catalogs which incorporate galaxy colors; these catalogs
are described in detail by \citet{gerke06b}.

To construct these new catalogs, we first measure the local overdensity of
each object in the \citet{yan04} mock catalogs using the full,
volume--limited catalog. Each galaxy is assigned a rest--frame $U-B$ color
drawn from the set of real DEEP2 galaxies at $0.8 < z < 1.0$ located in a
corresponding bin of local galaxy overdensity and absolute $B$--band
magnitude, employing the environment measurements of
\citet{cooper06a}. This produces mock catalogs which reproduce the observed
relationships between galaxy color, luminosity, and environment in the
DEEP2 data at $0.8 < z < 1.0$, but do not allow those relationships to
evolve with redshift unless the ``true'' distribution of galaxy
overdensities evolves. After galaxy colors are assigned, apparent $R$
magnitudes are determined using the K--correction methods applied for DEEP2
\citep{willmer06}, so that red galaxies will be lost at the same luminosity
at a given redshift as in the data. We then apply the standard DEEP2
target--selection and slitmask--making procedures \citep{davis03, faber06} to
these mock catalogs, so that we may directly determine the effects of DEEP2
target--selection algorithms on observed trends.

These mock catalogs should not be a perfect representation of reality, as
uncertainties in the observed environments will cause some objects to be
assigned the color of a galaxy that is actually in a higher-- or
lower--density environment than was measured; however, this effect is small
(i.e., environment measurement errors at $z = 0.8-1$ are significantly
smaller than the bin sizes). Nevertheless, they provide a robust test of
whether we may falsely observe an evolution in the relationship between
galaxy color and environment when in fact none exists.  They also allow us
to directly explore what effects the increased errors in environment
measurements at higher redshifts (where samples grow dilute) may have on
our results. We discuss these tests in \S 3.2.

\section{Results}

\subsection{The Color--Density Relation at $0.4 < z < 1.35$}

\begin{figure*}[h!]
\centering
\includegraphics[scale=0.6]{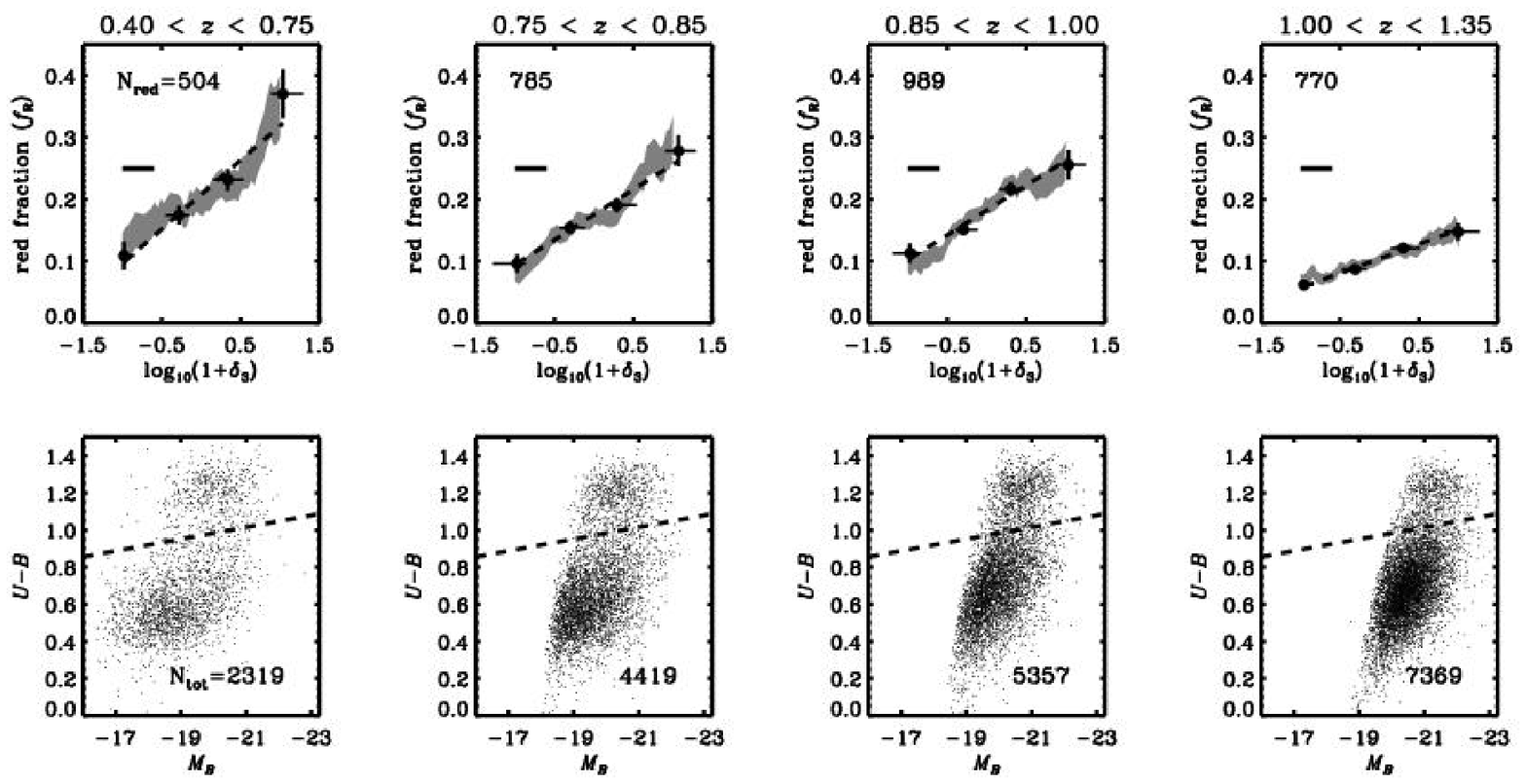}
\caption{\emph{Top Row:} Using the full galaxy sample (Sample A), we plot
  the red fraction as a function of overdensity, $\log_{10}(1 + \delta_3)$,
  in four distinct redshift ranges (redshift limits indicated at the top of
  each column). The circular points give the red fraction as function of
  the median overdensity, computed in four distinct bins of $\log_{10}(1 +
  \delta_3)$. The horizontal error bars run from the twenty--fifth
  percentile to the seventy--fifth percentile of the overdensity
  distribution in each bin. The vertical error bars give the 1--$\sigma$
  uncertainty on the red fraction within each overdensity bin, given by
  Poisson statistics. The grey shaded region in each panel shows the
  1--$\sigma$ range of the red fraction as a function of median overdensity
  in a sliding bin with width given by the horizontal line segment in the
  upper left of each plot $(\Delta \log_{10}{(1 + \delta_3)} = 0.2)$. The
  number in the upper left corner denotes the total number of red galaxies
  in the particular redshift interval. The dashed line in each panel shows
  a linear--regression fit to the four data points, with coefficients given
  in Table \ref{fits_tab}. \emph{Bottom Row:} We plot the rest--frame
  color--magnitude relation for all objects in each redshift bin. The
  division between the red sequence and the blue cloud is given by the
  dashed line, following the relation of \citet{willmer06} (cf.\ Equation
  \ref{willmer_cut}). The total number of galaxies (both blue and red) in
  each $z$ bin is enumerated in the bottom right corner and given in Table
  \ref{fits_tab}.}
\label{redfrac}
\end{figure*}

Figure \ref{redfrac} shows the relationship between the red fraction
$(f_{\rm R})$ and local galaxy overdensity for the full data set (Sample A)
in four distinct redshift bins spanning $0.4 < z < 1.35$. We find that the
red fraction exhibits a strong dependence on local environment in each
redshift bin, such that $f_{\rm R}$ is higher in regions of greater
overdensity. Fitting a linear relation to the trends in Figure
\ref{redfrac}, we find that the slope of this color--environment correlation
evolves with $z$, with the relative fraction of red galaxies in dense
environments decreasing with lookback time (cf.\ Table \ref{fits_tab}). Over
the full redshift range probed by the DEEP2 sample, the measured slope
decreases from ${\rm d}f_{\rm R} / {\rm d} \log_{10}(1 + \delta_3) = 0.111$
for $0.4 < z < 0.75$ to ${\rm d}f_{\rm R} / {\rm d} \log_{10}(1 + \delta_3)
= 0.046$ for $1.0 < z < 1.35$, with the difference being significant at a
greater than $3$--$\sigma$ level.

The vertical error bars in Figure \ref{redfrac} indicate the Poissonian
uncertainty in each point. Error estimates based on bootstrap and jackknife
resampling amongst the 10 DEEP2 pointings used yielded comparable
uncertainties (within 10--20\%), suggesting that sample (or ``cosmic'')
variance is not the dominant source of uncertainty; while cosmic variance
will influence the distribution of environments in a given redshift bin, it
should not strongly affect the relationship between galaxy color and
density in a given environment bin. Refer to \S 4.1 for further
investigation of the role of cosmic variance in this analysis.

\begin{figure*}[h!]
\centering
\includegraphics[scale=0.6]{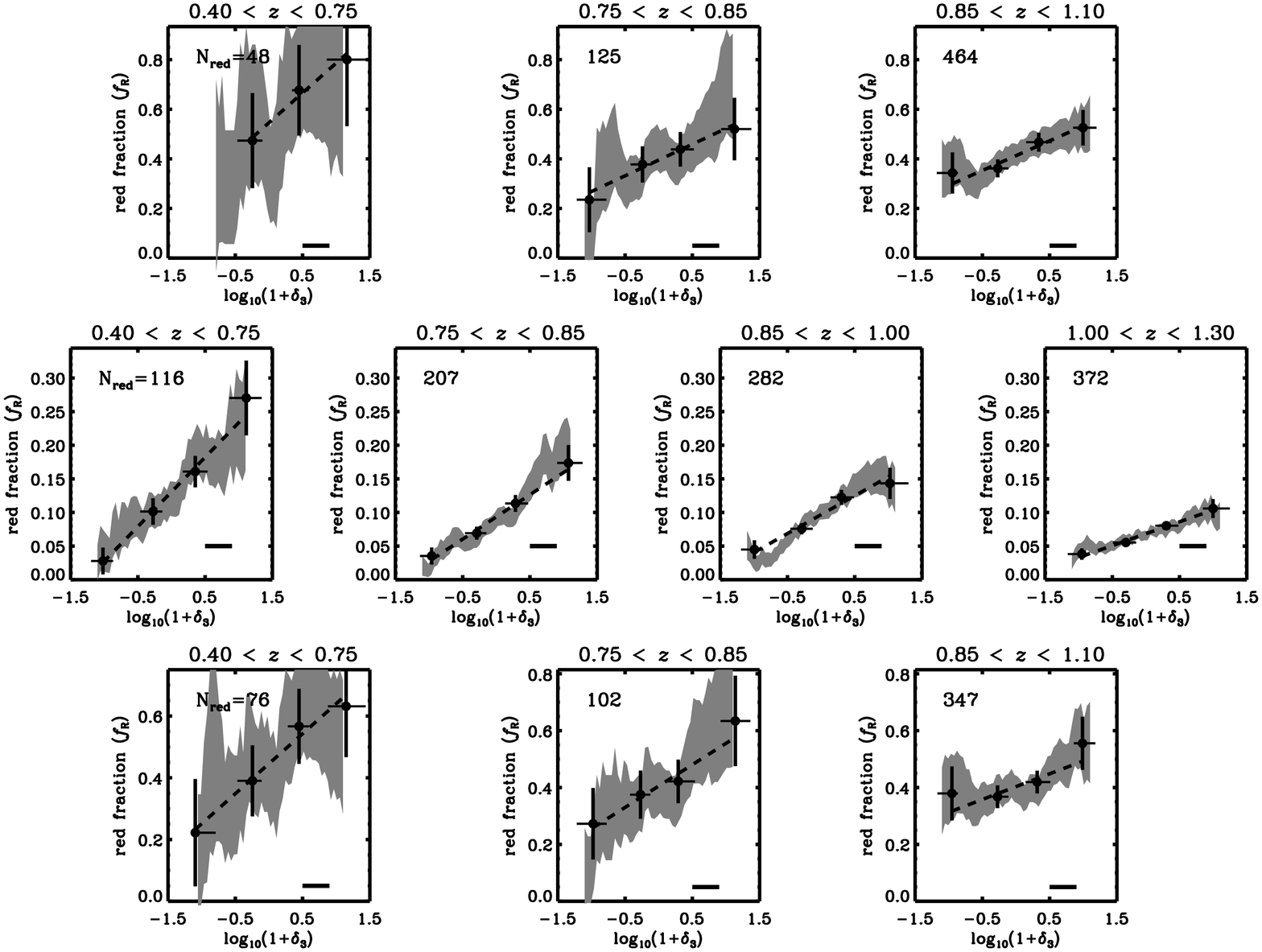}
\caption{We plot the red fraction as a function of overdensity,
  $\log_{10}(1 + \delta_3)$, in distinct redshift ranges for the remaining
  three galaxy samples identified in \S 2.3 and summarized in Table
  \ref{sample_descript_tab} --- Samples B (\emph{Top Row}), C (\emph{Middle
    Row}), and D (\emph{Bottom Row}). The circular points give the red
  fraction as function of the median overdensity, computed in distinct bins
  of $\log_{10}(1 + \delta_3)$. The horizontal error bars run from the
  twenty--fifth percentile to the seventy--fifth percentile of the
  overdensity distribution in each bin. The vertical error bars give the
  1--$\sigma$ uncertainty on the red fraction within each overdensity bin,
  given by Poisson statistics. The grey shaded region in each panel shows
  the 1--$\sigma$ range of the red fraction as a function of median
  overdensity in a sliding bin with width given by the horizontal line
  segment in the lower right of each plot $(\Delta \log_{10}{(1 +
    \delta_3)} = 0.2)$. The number in the upper left corner denotes the
  total number of red galaxies in the particular redshift interval. The
  dashed line in each panel shows a linear--regression fit to the data
  points, with coefficients of the fit given in Table \ref{fits_tab}. For
  the each sample, the baseline value of $f_{\rm R}$ varies significantly
  as the different sample selection criteria include different portions of
  color--magnitude space.}
\label{redfrac_BCD}
\end{figure*}

All of the samples described in \S 2.3 yield similar results for the
evolution of the color--density relation (cf.\ Fig.\ \ref{redfrac_BCD} and
Table \ref{fits_tab}). While overall $f_{\rm R}$ values by overdensity
differ when either a color--dependent or a color--independent $M_B$ limit
is applied to the full sample, in every case the color-environment relation
weakens, but still is present, at higher $z$. Due to their small sample
sizes, the effect is not statistically sigificant for Samples B and D,
however.

\begin{deluxetable*}{c c c c c c c}
\tablewidth{0pt}
\tablecolumns{5}
\tablecaption{\label{var_tab} Linear Fits to the $f_{\rm
R}$-overdensity Relation}
\tablehead{ & $N_{\rm red}$ &  $N_{\rm tot}$ & 
$a_0$ (slope) & $a_1$ ($y$--intercept) & $\sigma_{\rm slope}$ & $\sigma_{y{\rm-int}}$ }
\startdata
Sample A& & & & \\
$0.4 < z < 0.75$ & 504 & 2319 & 0.111 & 0.208 & 0.018 & 0.010 \\
$0.75 < z < 0.85$ & 785 & 4419 & 0.081 & 0.175 & 0.012 & 0.007 \\
$0.85 < z < 1.0$ & 989 & 5357 & 0.080 & 0.182 & 0.012 & 0.006 \\
$1.0 < z < 1.35$ & 770 & 7369 & 0.047 & 0.104 & 0.007 & 0.004 \\
\hline \\
Sample B & & & & \\
$0.4 < z < 0.75$ & 48 & 76 & 0.240 & 0.546 & 0.230 & 0.140 \\
$0.75 < z < 0.85$ & 125 & 297 & 0.127 & 0.395 & 0.076 & 0.044 \\
$0.85 < z < 1.1$ & 464 & 1081 & 0.120 & 0.413 & 0.047 & 0.024 \\
\hline \\
Sample C & & & & \\
$0.4 < z < 0.75$ & 116 & 813 & 0.103 & 0.131 & 0.019 & 0.013 \\
$0.75 < z < 0.85$ & 207 & 2143 & 0.066 & 0.093 & 0.012 & 0.007 \\
$0.85 < z < 1.0$ & 282 & 2868 & 0.056 & 0.097 & 0.011 & 0.006 \\
$1.0 < z < 1.3$ & 372 & 5368 & 0.035 & 0.068 & 0.007 & 0.004 \\
\hline \\
Sample D & & & & \\
$0.4 < z < 0.75$ & 76 & 148 & 0.194 & 0.446 & 0.098 & 0.069 \\
$0.75 < z < 0.85$ & 102 & 237 & 0.150 & 0.405 & 0.086 & 0.049 \\
$0.85 < z < 1.1$ & 347 & 844 & 0.090 & 0.403 & 0.056 & 0.026 \\
\enddata

\tablecomments{We list the coefficients and 1--$\sigma$ uncertainties for
  the parameters of the linear-regression fits to the red fraction versus
  median overdensity relation given by $f_{\rm R} = a_0 \cdot \log_{10}(1 +
  \delta_3) + a_1$ (cf.\ Fig.\ \ref{redfrac} and Fig.\ \ref{redfrac_BCD}) in
  each $z$ bin employed for all four galaxy samples used. We also give the
  number of red--sequence members $(N_{\rm red})$ along with the total
  number of galaxies $(N_{\rm tot})$ in each redshift bin for that
  sample. For details regarding the various galaxy samples, refer to \S 2.3
  and Table \ref{sample_descript_tab}.}
\label{fits_tab}
\end{deluxetable*}

\begin{figure}[h!]
\centering
\plotone{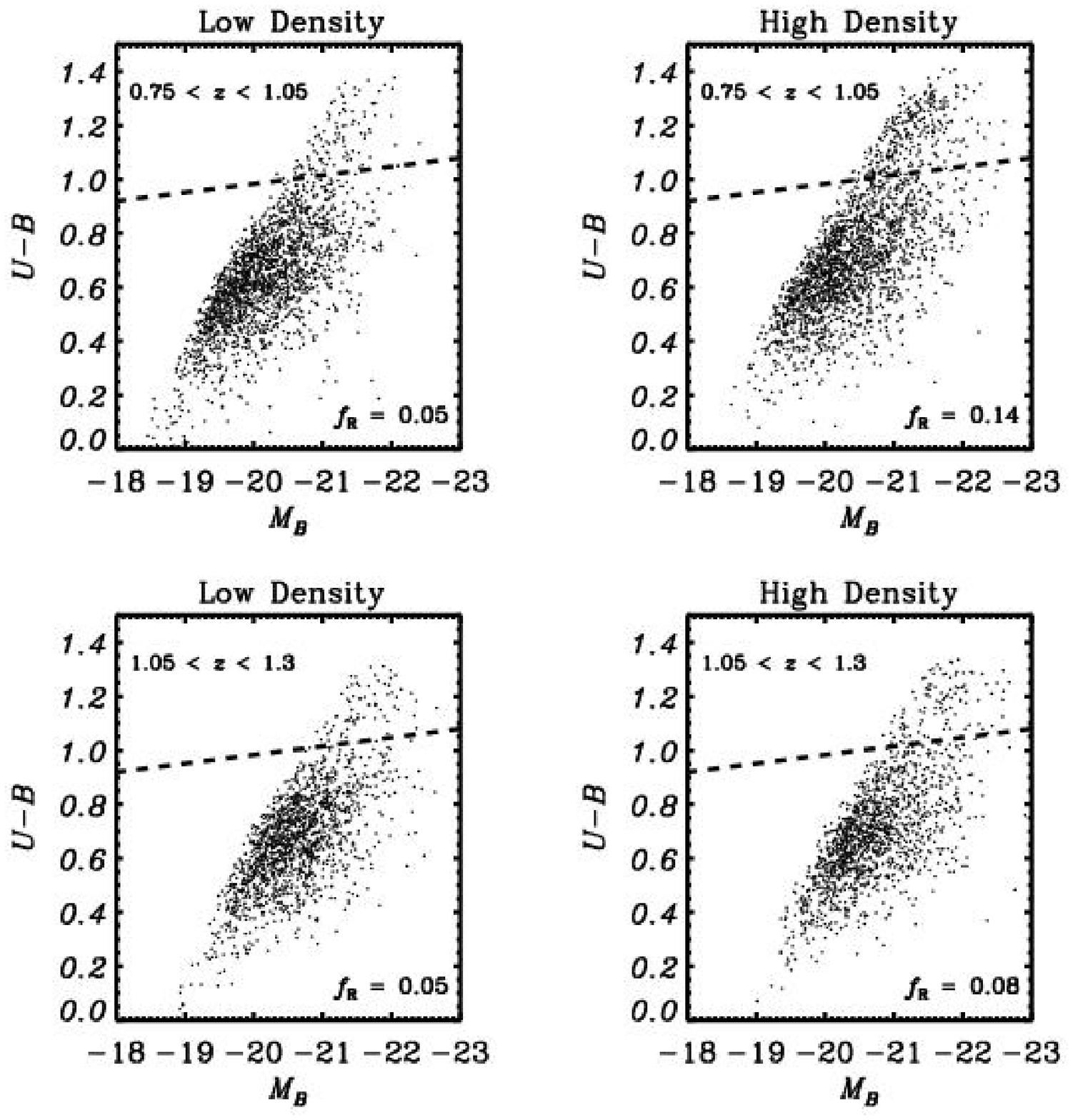}
\caption{We plot the color--magnitude relation for DEEP2 galaxies in
  low--density and high--density environments at low and high redshift. Here
  we use ``low'' and ``high'' density to refer to the extreme thirds of the
  overall overdensity distribution within Sample C. The dashed line in
  each plot denotes the division between the red sequence and blue cloud as
  given in Equation \ref{willmer_cut}. At the bottom right corner of each
  plot, the red fraction $(f_{\rm R})$ for that color--magnitude diagram is
  given. For each row, the two panels contain equal numbers of objects. The
  greater prevalence of red objects in high--density regions, particularly
  at lower redshifts, is readily apparent to the eye.}
\label{cmd_4panel}
\end{figure}

To further illustrate the evolution in the color--density relation with
$z$, we present color--magnitude diagrams in Figure \ref{cmd_4panel}
divided by local environment and redshift. In this figure, the relationship
between red fraction $(f_{\rm R})$ and galaxy overdensity is made directly
discernable to the eye; in high--density regions, there is an increased
proportion of sources on the red sequence relative to that observed in
low--density environments.

\begin{figure*}[h!]
\centering
\plotone{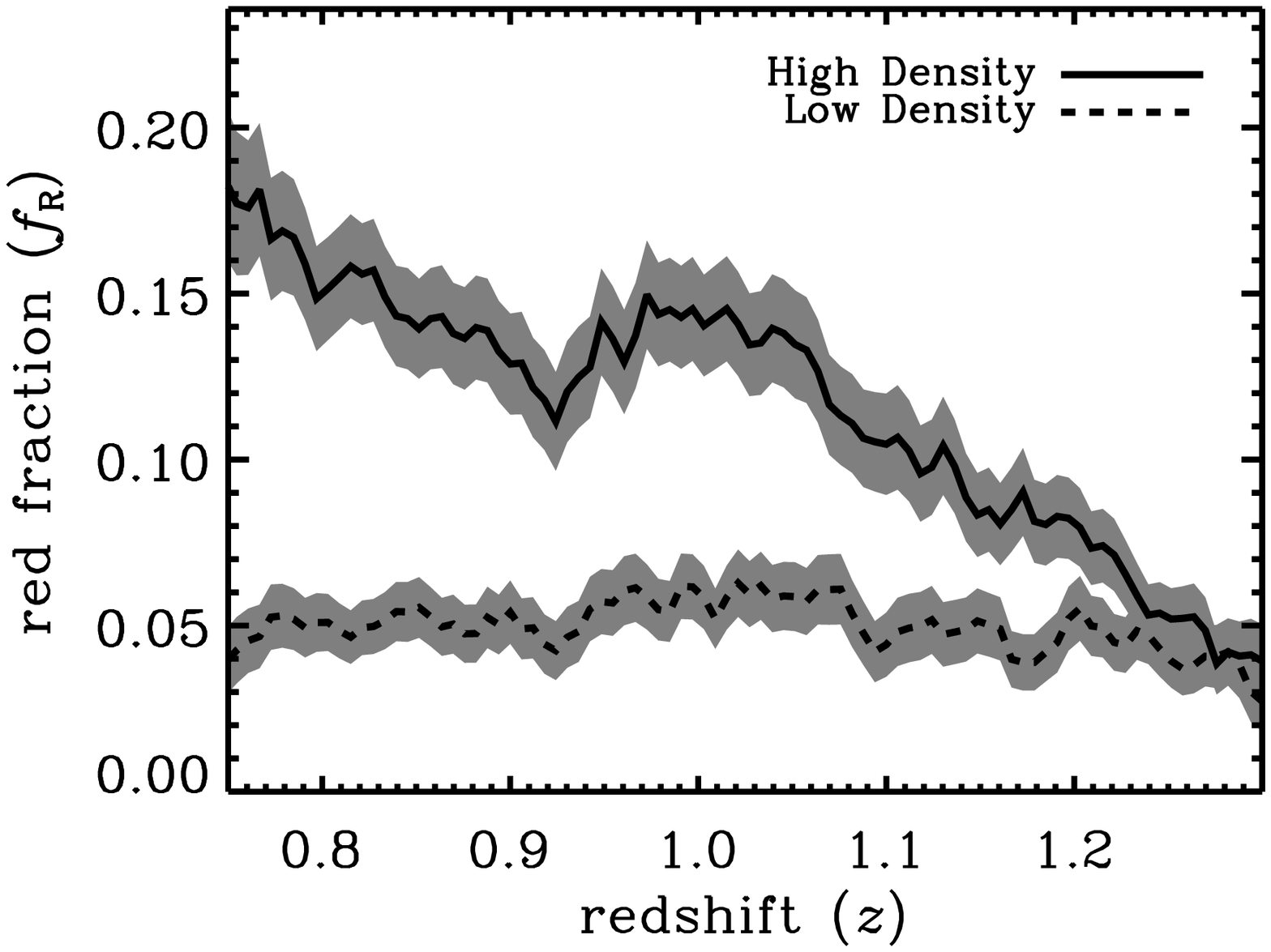}
\caption{For galaxies in Sample C within high--density (\emph{solid line})
  and low--density (\emph{dashed line}) environments, we plot the red
  fraction $(f_{\rm R})$ as a function of redshift for galaxies in sliding
  bins of $\Delta z = 0.1$. The high-- and low--density samples are selected
  according to the extreme thirds of the local overdensity $(1 + \delta_3)$
  distribution in the given $z$ bin. The grey shaded regions give the
  1--$\sigma$ range of the red fractions in each density regime.  The
  color--density relation evolves significantly over the redshift range
  $0.75 < z < 1.3$, with the red fraction in high--density environments
  dropping continuously as $z$ increases, while in low--density regions it
  remains relatively constant at all redshifts probed. At $z \sim 1.3$, the
  red fraction shows no dependence on environment within the measurement
  errors.}
\label{redfrac_withz}
\end{figure*}

While the results shown in Figure \ref{redfrac} show a clear signature of
evolution in the color--density relation over the redshift range $0.4 < z <
1.35$, relatively large redshift bins are used, thereby coarsely sampling
the redshift domain. To study the evolution of the $f_{\rm R}$--overdensity
relationship with $z$ in more detail, we divide Sample C into thirds
according to overdensity and compute the red fraction as a function of
redshift in a sliding bin of width $\Delta z = 0.1$ for the galaxies in the
high--density $(\log_{10}{(1 + \delta_3)} \gtrsim 0.25)$ and low--density
$(\log_{10}{(1 + \delta_3)} \lesssim -0.2)$ extremes of the overdensity
distribution, with results shown in Figure \ref{redfrac_withz}. Here, the
high--density and low--density thirds are defined with respect to the
overdensity distribution in each $z$ bin; however, dividing the galaxy
sample into high-- and low--density regimes according to the overdensity
distribution in the entire redshift range $(0.75 < z < 1.3)$ probed yields
consistent results. As detailed in \S 2.3, sample C is defined so as to be
complete with respect to $M_{B}^{*}$ across the full redshift range, $0.4 <
z < 1.3$, establishing a sample uniformly selected at all redshifts
studied.

As illustrated in Fig.\ \ref{redfrac_withz}, the color--density relation
shows a continuous evolution with redshift from $z \sim 0.75$ to $z \sim
1.3$, such that at redshifts approaching $z \sim 1.3$ the red fraction in
low-- and high--density regions are statistically consistent with each
other. Extrapolating linear regression fits to the $f_{\rm R}(z)$ relations
in high--density and low--density environments, we find convergence at a
redshift of $z = 1.32$. We investigate this convergenece in more detail in
the following subsections (\S 3.1.2 and \S 3.1.3).

In our analyses, we have attempted to carefully account for selection
effects due to the design parameters of the DEEP2 survey. For all galaxy
samples but Sample A, selection effects related to the survey's magnitude
limit and $R$--band selection, which bias the sample against faint and/or
red galaxies at higher redshifts, shoud be minimal. The agreement amongst
the highly disparate samples indicates that the results presented above are
robust to such effects.

These results are similar to the findings of \citet{gerke06b}, which
investigates the evolution of the blue fraction among group and field
galaxies at $0.7 < z < 1.3$ in DEEP2. That work employed a very different
technique for measuring galaxy environment and studied a somewhat different
set of galaxy subsamples. The measurements of the evolution in the blue
fraction presented by \citet{gerke06b}, however, are significantly noisier
than those presented in this study and could be susceptible to biases if
the group--finding algorithm used performs differently at different $z$. We
note that while that work uses a different methodology for dividing
galaxies into environment bins (identifying members of galaxy groups rather
than with a continuous measure of local galaxy density), both the
\citet{gerke06b} results and those presented here are derived from
subsamples of the DEEP2 survey; hence any systematics affecting one may
also affect the other.

\subsubsection{Effect of Possible Evolution in the Color of the Bimodality}

While the position of the bimodality in $U-B$ versus $M_B$ color--magnitude
space shows no significant dependence on redshift within the DEEP2 data,
when studying the red fraction as a function of redshift, we must consider
the possibility that there is evolution in the color of the red--sequence
population and thus in the position of the color bimodality. If the
location of the color bimodality evolves with reshift, then the measured
evolution in the red fraction in high--density environments could simply
result from galaxies moving (e.g., via passive evolution) across our
non--evolving color division (cf.\ Equation \ref{willmer_cut}). However, any
affect related to possible evolution in the location of the $U-B$
bimodality would be minimized by the differential nature of our
measurements. That is, the location of the color bimodality does not appear
to depend on environment, and thus any evolution in the red fraction for
the galaxy sample in high--density regions should be mirrored in the $f_{\rm
  R}(z)$ for the sample of galaxies in intermediate-- and low--density
environments.

Furthermore, given that passive evolution will cause galaxies to move
redward as they evolve and that the $U-B$ color division in Equation
\ref{willmer_cut} is defined by a galaxy sample at $z \sim 0.8$
\citep{willmer06, vandokkum01},
a non--evolving color division will \emph{not} increase the measured
evolution in the color--density relation due to an interloper population of
blue galaxies on the red sequence at $z > 1$. Instead, a non--evolving cut
would create a more stringent selection criterion at higher redshifts,
limiting the number of galaxies residing in the trough between the red
sequence and blue cloud (also called the ``green valley'') that are counted
as members of the red sequence, thereby actually producing weaker evolution
in the color--density relation.

In the spirit of thoroughness, however, we test the effect of using an
evolving color division on our results. Allowing for passive evolution in
the location of the color bimodality of $\sim 0.15$ magnitudes (in $U-B$)
per unit $z$ \citep[e.g.,][]{vandokkum01, blanton06a}, such that the
division between red and blue galaxies moves blueward with lookback time,
we find no significant change in our results. That is, employing a
redshift--dependent division between the blue cloud and red sequence, the
relative red fraction in high-- and low--density environments still shows
strong evolution, consistent with no color--density relation at $z \sim
1.35$.



\subsubsection{Effect of Redshift Incompleteness}

In analyzing the evolution of the red fraction $(f_{\rm R})$ as a function
of redshift, we must consider the possible impact of redshift--dependent
selection effects in the DEEP2 survey. In particular, due to the finite
amount of slit--length real estate available on DEIMOS slitmasks and the
finite amount of observing time dedicated to the project, the DEEP2 survey
only spectroscopically observes $\sim \! 60\%$ of the galaxies that meet the
survey's target--selection criteria. Furthermore, about $30\%$ of those
galaxies targeted for spectroscopy fail to yield a high--quality (quality $Q
= 3,4$) redshift.

Initial follow--up observations of sources for which DEEP2 fails to measure
a redshift indicate that roughly half of all failures (i.e., $\sim \! 15\%$
of DEEP2 targets) are at redshifts beyond the range probed by DEEP2 (i.e.,
$z \gtrsim 1.4$; C.\ Steidel, private communication). Additionally,
redshift failures may result from poor observing conditions, data
reductions errors, or instrumental effects. Clearly important for this work
is the fact that the DEEP2 redshift failure rate is correlated with
observed galaxy color and magnitude, thus possibly introducing bogus trends
in the measured $f_{\rm R}(z)$ relation.

To understand the degree to which our results are impacted by redshift
incompleteness, we employ the weighting scheme presented by
\citet{willmer06}. The derived weights account for variations in both
redshift incompleteness and targeting rate as a function of apparent $R$
magnitude and apparent $R-I$ and $B-R$ colors, assuming that the redshift
distribution for red $(R-I > 1.03)$ failures is identical to that of the
observed (quality $Q = 3,4$) red galaxy sample and that the blue $(R-I <
1.03)$ failures sit beyond the redshift range of the survey \citep[defined
as the ``optimal'' model in][]{willmer06}. If we use these weights to
compute $f_{\rm R}(z)$ for Sample C (as in Fig.\ \ref{redfrac_withz}), all
changes in the measured trends with redshift are well within the
measurement uncertainties. Thus, biases related to redshift incompleteness
should not introduce spurious results in our analysis. With and without
weighting for incompleteness, we find a convergence of the red fraction in
high-- and low--density environments at $z \gtrsim 1.3$; extrapolating linear
regression fits to the $f_{\rm R}(z)$ relations in high--density and
low--density environments, we find convergence at a redshift of $z = 1.33$
when weighting for incompleteness versus a convergence at $z = 1.32$
without weighting.

\subsubsection{Tests with Mock Galaxy Catalogs}

The mock galaxy catalogs described in \S 2.4 allow us to further test the
degree to which our observations of the color--density relation may be
subject to systematic effects within DEEP2 (e.g., due to increased errors
in environment measures as the sampling density drops with $z$ or due to
the expected changes in the magnitude of peculiar velocities with $z$
compared to the 1000 ${\rm km}/{\rm s}$ window used). These catalogs should
exhibit no evolution in the color--density relation when measured with
perfect information, by construction. Therefore, when we apply our
measurement techniques to subsets of the mock catalogs which emulate the
observed samples, any apparent evolution in the color--density relation will
generally indicate the presence of observational biases.

\begin{figure}[h!]
\centering
\plotone{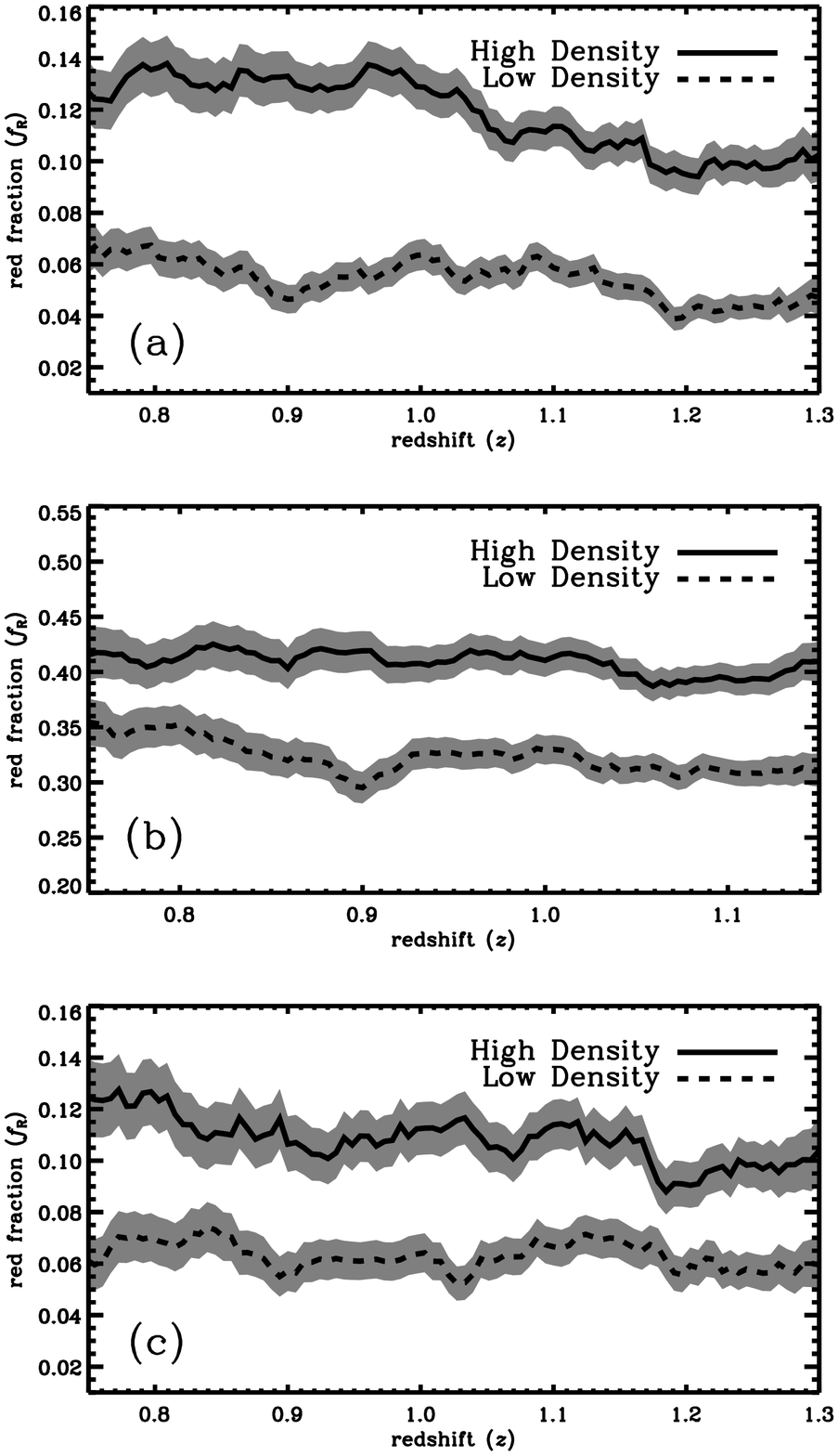}
\caption{For the mock galaxy samples, we plot the red fraction $(f_{\rm
    R})$ as a function of redshift in high--density (solid line) and
  low--density (dashed line) environments where the high--density and
  low--density samples are selected as the extreme thirds of the
  environment distribution for the given galaxy sample. The shaded grey
  regions in each plot trace the 1--$\sigma$ range of the red fractions in
  each density regime. Both plots (a) and (b) utilize ``true'' environment
  measures defined using the entire volume--limited mock catalog. However,
  (a) shows $f_{\rm R}(z)$ for a DEEP2--like ``observed'' sample, while (b)
  gives the same relation for a sample selected according to a fixed (that
  is, redshift--independent) $M_{B} < -20$ absolute--magnitude limit. Like
  (a), panel (c) uses a DEEP2--like ``observed'' sample. In contrast,
  though, panel (c) employs environments estimated according to the
  overdensity $(1 + \delta_3)$ as measured in the ``observed'' sample
  (i.e., using redshift--space galaxy positions and the DEEP2--like galaxy
  sample). The differences between (a) and (c) are consistent with a
  scenario where noise in environment measures causes the color--density
  relation at high redshift to be smeared out. However, such observational
  selection effects cannot alone explain the convergence of $f_{\rm R}$
  that we observe at $z \sim 1.3$ in the DEEP2 data (cf.\ Fig.\
  \ref{redfrac_withz}). For more details regarding the specific sample
  selection criteria and environment measures, refer to the text of \S
  3.1.3.}
\label{chromo_plots}
\end{figure}

There is a notable exception to this. It is possible to select a subsample
from the volume--limited mock galaxy catalog (i.e., the catalog before
applying DEEP2 target--selection criteria or slitmask--making algorithms)
using a color--dependent, absolute--magnitude limit which evolves with $z$, as
for our Sample C above; however, for the mock catalogs, $Q = -1$ is the
appropriate evolution, as that value was used in their construction, rather
than $Q = -1.37$ as used for Sample C. We find that if we use the full
volume--limited mock catalog to determine local overdensities in real--space,
there is in fact evolution in the color--density relation for such a sample
(cf.\ Figure \ref{chromo_plots}a).

This evolution does not reflect an observational selection effect,
however. Instead, it is a result of applying an absolute--magnitude cut
which changes with $z$ to mock catalogs which were constructed assuming
rest--frame color depends on $M_B$ in a way which is independent of $z$ (cf.\
\S 2.4). Because $M_{B}^*$ was brighter at higher $z$, the galaxies
included by DEEP2 tend to be brighter at higher $z$; and as may be seen in
Fig.\ \ref{sample_select} and Fig.\ \ref{cmd_4panel}, brighter galaxies are
more likely to be red. The color--dependent, absolute--magnitude limit
exaggerates this effect. If instead we apply a color-- and
redshift--independent absolute--magnitude limit to the volume--limited mock
sample, the mock catalogs show considerably weaker color--density evolution,
as we would expect (cf.\ Fig.\ \ref{chromo_plots}b). For panels (a) and (b)
of Figure \ref{chromo_plots}, we determine the ``true'' local galaxy
environment according to the $7^{\rm th}$--nearest--neighbor surface density
(converted into an overdensity, $(1 + \delta_7)$) using the real--space
galaxy positions and the full volume--limited galaxy population.

Although the mock catalogs exhibit a modest evolution in the color--density
relation for samples with an absolute--magnitude limit that shifts with
$M^*$ (like our Sample C), they still may be used to determine the net
effects of DEEP2 sample selection on the observed color--density
relation. We do this by comparing $f_{\rm R}(z)$ in two different mock
catalog samples. For the first, we do not apply DEEP2 target--selection and
slitmask--making criteria, but only the color--dependent absolute--magnitude
limit given by Equation \ref{magcut} with $Q=-1$, and we use the best
possible measure of environment in the catalogs \citep[based upon the
real--space overdensity of galaxies in the full, volume--limited catalog, as
employed by][]{cooper05} to divide the sample into environment bins; see
Fig.\ \ref{chromo_plots}a. For the second, we apply the DEEP2
target--selection and slitmask design algorithm to these mock catalogs, and
use the overdensity measured from this ``observed'' sample to partition the
sample; results for this sample are shown in Fig. \ref{chromo_plots}c.

We find that in fact the DEEP2--like sample exhibits both a smaller gap
between $f_{\rm R}(z)$ for the extreme environment bins and a stronger
apparent evolution over $0.75 < z < 1.3$. This is consistent with a
scenario where noise in environment measures causes objects to cross the
boundary between the intermediate environment bin and the extremes, thereby
artificially increasing the red fraction for the low--density sample while
decreasing $f_{\rm R}$ for the high--density sample. This effect should
become worse at higher $z$, as the sampling density decreases there,
leading to greater environment errors. Therefore, though we observe
convergence of between the red fraction in low-- and high--density
environments at $z = 1.32$, in actuality convergence should not occur until
a somewhat higher redshift.

However, this test shows that observational selection effects cannot
explain the convergence in $f_{\rm R}$ that we observe in high--density and
low--density regions at $z \sim 1.3$. Extrapolating linear--regression fits
to the $f_{\rm R}(z)$ relations in high--density and low--density
environments, we find convergence at $z=2.50$ if we measure environment
with the full sample (Fig.\ \ref{chromo_plots}a), or $z = 2.29$ after we
apply DEEP2 target selection procedures (Fig.\ \ref{chromo_plots}c). The
mock catalogs provide clear evidence that the null hypothesis of no
evolution in the color--magnitude relation over $0.75 < z < 1.3$ cannot
hold.

\subsection{Red Galaxies in Low--Density Environments}

The relations presented in Figure \ref{redfrac} also show evidence of a
trend with $z$ in the normalization of the $f_{\rm R}$--overdensity
relation, such that the total fraction of red--sequence galaxies averaged
over all densities decreases with $z$. This effect is dominated by two
systematic trends in the data: (1) from $z \sim 0$ to $z \sim 1$, the
number density of red galaxies in the universe decreases by a factor of
$\sim \! 2$--3 \citep[e.g.,][]{bell04, faber05}, and (2) at redshifts
beyond $z \sim 1.1$, the red galaxy fraction decreases precipitously in the
DEEP2 sample, due to the survey's $R$--band magnitude limit
\citep{cooper06a}. This latter effect is quite apparent in the final
redshift bin in Figure \ref{redfrac}, where the normalization of the
observed red fraction--overdensity relation drops significantly; the other
volume--limited samples studied here should not be affected by this, however
(cf.\ Fig.\ \ref{redfrac_BCD}).

In spite of these redshift--dependent effects, we still find that at all
redshifts probed in this paper some fraction of red galaxies populate very
underdense environments. While uncertainties in the environment
measurements will lead to some galaxies at intermediate densities being
scattered into the low--density third of the overdensity distribution, upper
limits on this effect show that it cannot account for all of the red
galaxies found in low--density environments at any redshift, as we now
show. We proceed by assuming that the observed $\log_{10}{(1 + \delta_3)}$
distribution of red galaxies matches the shape of the true distribution, an
assumption that holds so long as the distribution is approximately linear
over scales comparable to measurement errors (as convolving a linear
function with a Gaussian leaves it unchanged). In actuality, e.g. if the
$\log_{10}{(1 + \delta_3)}$ distribution has some cutoff value at low
overdensity, the observed distribution will have more objects below this
cutoff than the true distribution; thus the observed distribution provides
an upper limit to the true contamination rate.

We calculate this limit from the expectation value of the number of red
objects that have true overdensities in the top two--thirds of galaxies but
measured overdensities in the lowest one--third, minus the number of objects
truly in the bottom one--third but measured in the top two--thirds. We assume
that each overdensity measurement may be represented by a Gaussian
distribution in $\log_{10}{(1 + \delta_3)}$, with mean given by the
measured value and standard deviation increasing linearly from $\sigma =
0.49\ {\rm dex}$ at $z=0.8$ to $0.59\ {\rm dex}$ at $z=1.2$, based upon
tests with the mock catalogs described in \S 2.4. We find that the
increasing uncertainties in environment measures combined with the
weakening in the strength of the color--density relation with lookback time
yield a net contamination rate that is roughly constant with redshift
$(\sim 35\%)$. Hence, the true red fraction in the extreme low--density
third of the sample is at least $3\%$ at all redshifts. Similar studies at
lower redshift \citep[e.g.,][]{balogh04b, yee05, croton05, martinez06} have
shown that the existence of red galaxies in low--density environments
persists to $z \sim 0$.

Clearly, a population of red galaxies in low--density environments exists at
$0.75 < z < 1.3$. However, galaxies may appear red either because they are
true red--sequence/early--type galaxies or because of the presence of
interstellar dust. Based on past studies \citep[e.g.,][]{lotz06}, we might
expect the former population to dominate the red sequence at lower redshift
and the latter at higher $z$. Are the red--sequence members in low--density
regions at low and high redshift comparable in terms of galaxy morphology?
Out of the less than 40 red galaxies in underdense environments identified
in DEEP2 for which {\it HST}/ACS imaging is available in the EGS
\citep{davis06}, an initial by--eye inspection indicates that the majority
of this population exhibits early--type morphologies, with the remainder
being reddened disk galaxies. While the sample imaged with {\it HST}/ACS is
small, we find no significant trends of morphology with redshift; for
example, dusty disk galaxies do not increasingly dominate the red sequence
in low--density environments at higher $z$. However, contamination by
late--type galaxies preferentially at higher $z$ cannot be excluded as a
possible explanation for the observational results presented here; an
analysis of galaxy morphologies by \citet{lotz06} concludes that late--type
galaxies comprise 30\% of the red sequence at $z \gtrsim 1$, though they
are only a minor contaminant at low redshift \citep[see
also][]{weiner05}. Such an increase in the contribution of dusty disks to
the red sequence at $z > 1$ could work to reduce the strength of an
existing color--density
relation.

Are members of the red sequence in underdense regions the result of passive
evolution, in which they consumed their gas supplies independent of
environmental effects, or are they fossil groups that result from the
merging of several smaller galaxies? Clearly, even in underdense
environments galaxy mergers are capable of triggering events that reduce
gas reservoirs and at least temporarily halt star formation as well as
disrupt galactic disks, yielding merger remnants with surface--brightness
profiles and density distributions similar to those of early--type galaxies
\citep{toomre72, navarro87, barnes92, hopkins05}. Similarly, passive
evolution when teamed with processes attributed to secular evolution such
as bar instabilities could also explain the existence of red,
morphologically early--type galaxies in low--density environments
\citep{zhang96}. In future work, we hope to explore the morphologies of red
galaxies in low--density regions within the EGS in more detail, with a goal
of differentiating between these two scenarios; galactic bulges formed by
bar instabilities tend to differ from those built via mergers in that the
former exhibit more disk--like properties such as flatter profiles and
residual bars or spiral structure \citep{kormendy05}.

\section{Discussion}

\subsection{Comparison with Related Studies}

From our study of 19,464 galaxies in the DEEP2 Galaxy Redshift Survey, we
conclude that the color--density relation observed in the local universe is
also seen at $z > 1$, with the fraction of red galaxies increasing with
local galaxy overdensity at essentially all epochs studied $(0.4 < z <
1.35)$ and over all environments probed (from voids to large groups). At
all redshifts, however, there still exists a population of red galaxies in
underdense environments. In addition, we find that the color--density
relation evolves with redshift, growing weaker with lookback time such that
at $z \gtrsim 1.3$ the relationship may be nonexistent within the range of
environments probed by the DEEP2 survey (i.e., not including massive
clusters). When viewed in conjunction with the results from studies of the
galaxy luminosity function at $0 < z < 1$ \citep{bell04, faber05}, our
findings provide direct evidence that 
the red sequence is built up preferentially in overdense environments
(i.e., galaxy groups, as DEEP2 does not sample rich clusters), thereby
producing the observed increase in the slope of the red fraction versus
overdensity relation at later time.

These results are in agreement with the general picture painted by studies
of the morphology--density relation in galaxy clusters at $0 < z < 1$. For
example, building upon the work of \citet{dressler97}, \citet{smith05} and
\citet{postman05} find that the fraction of early--type galaxies increases
steadily with density in cluster environments out to $z \sim 1$, with the
strength of the correlation weaker at $z \sim 1$ than in local
samples. Corresponding work by \citet{vandokkum00} also finds significant
evolution in the morphology--density relation within massive clusters at $z
< 1$, with the early--type fraction observed to steadily decline with
increasing redshift.

Using data from the Canada--France--Hawaii Telescope Legacy Survey
(CFHTLS), \citet{nuijten05} similarly find that over the redshift range $0
< z < 1$ both the red fraction and early--type fraction increase in
high--density regions with decreasing redshift. While the fraction of
galaxies with early--type morphologies in the \citet{nuijten05} sample is
constant with $z$ in low--density environments, they find that the red
fraction $(u-g > 1)$ steadily increases in the ``field'' from $z \sim 0.8$
to $z \sim 0$, in constrast to our results which show no evolution in the
red fraction in regions of low galaxy density at $0.75 < z < 1.3$. The
differences between our results and those of \citet{nuijten05} could result
from the use of photometric redshifts to determine environments in that
work; as shown by \citet{cooper05}, photometric redshifts cannot cleanly
discriminate the environments of galaxies, as even the smallest
photometric--redshift errors achieved are much larger than galaxy
correlation lengths ($< 5 h^{-1}$ comoving Mpc versus $> 15 h^{-1}$
comoving Mpc for $\sigma_{z}=0.01$).

\begin{figure*}[h!]
\centering
\includegraphics[scale=0.6]{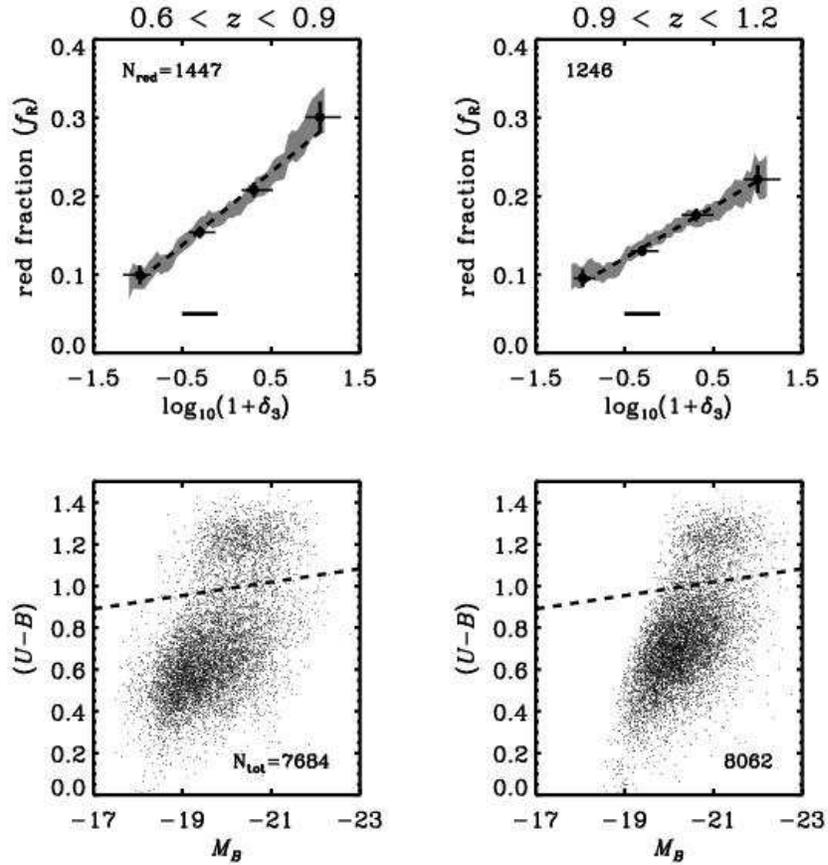}
\caption{As in Figure \ref{redfrac}, except we plot the red fraction as a
  function of overdensity and the rest--frame color--magnitude distribution
  in only two distinct redshift ranges for the full galaxy sample (Sample
  A). Here, the redshift bins are selected to match those of
  \citet{cucciati06}. In comparison to Figure 6 of \citet{cucciati06}, we
  see a highly significant environment trend where they do not.}
\label{direct_comp}
\end{figure*}

Our findings also reproduce the general trend found by \citet{cucciati06}
based on the VVDS survey; they, too, find that the color--density relation
weakens with redshift over the range $0 < z < 1.5$. However, within their
study the fraction of galaxies on the red sequence shows no significant
dependence on overdensity at $z > 0.9$. In contrast, we find a highly
significant relationship between red fraction and environment at $z \sim
1$, even for highly differing subsamples; only at $z \gtrsim 1.3$ are our
results consistent with density independence, as seen in Fig.\
\ref{redfrac_withz}. When binning the DEEP2 data in the same redshift
ranges as that of \citet{cucciati06}, the differences between the DEEP2 and
VVDS results are 
readily apparent \citep[see Figure \ref{direct_comp} of this work and
Figure 6 of][]{cucciati06}. Given this apparent contradiction for $z
\gtrsim 0.9$ and that the work of \citet{cucciati06} employs the most
analogous data set to that presented here, a more detailed examination is
required.

First of all, because rest--frame galaxy colors (which depend on photometry
and coarsely on redshift) are almost entirely independent of the
environment measurements (which depend upon angular positions and
high--precision redshifts), we consider it unlikely that the highly
significant correlation between red fraction and environment, which we find
at $z \sim 1$, is false. It persists in all four samples considered here,
which should differ from each other by more than sample B should differ
from the VVDS volume--limited sample. Since there are few galaxies in the
gap between the red and blue populations, modest differences in the
definition of $f_{\rm R}$ have minimal effect, as well, so this is unlikely
to explain any differences.


In an effort to conduct a more direct comparison between the VVDS and DEEP2
results, we attempt to replicate as closely as possible those VVDS
subsamples for which we can establish a volume--complete DEEP2 analogue and
apply to them the same red--fraction definition as employed by
\citet{cucciati06}. While in this paper we generally define the red
fraction, $f_{\rm R}$, according to a color division in $U-B$ versus $M_B$
color--magnitude space, \citet{cucciati06} utilize a
luminosity--independent selection in $u^{*}-g'$. Using the CFHT/Megacam
$u^{*}$ and $g'$ filter response, quantum efficiency, telescope throughput,
and atmospheric extinction
estimates\footnote{http://www.cfht.hawaii.edu/Instruments/Filters}, the
K--correction code (\emph{kcorrect} version v4\_1\_2) of \citet{blanton03},
and our CFHT $B,R,I$ photometry \citep{coil04b}, we compute the rest--frame
$u^{*}-g'$ color for each galaxy in the DEEP2 spectroscopic
sample. \citet{cucciati06} divide samples according to Johnson/Cousins
$B$--band absolute magnitude in the AB system, taking $h = 1$; this is
identical to the $M_B$ used throughout this paper (cf.\ \S 2.2). We are
therefore able to place the DEEP2 galaxies in the same color--magnitude
space as that of \citet{cucciati06} and define ``red'' according to the
VVDS definition ($u^{*}-g' \ge 1.1$).

In order to match the VVDS samples as closely as possible, we have
constructed DEEP2 samples covering identical redshift regimes ($0.6 < z <
0.9$ and $0.9 < z < 1.2$) and absolute--magnitude limits ($M_B \le -20,
-20.5, -21$) as VVDS subsamples studied by \citet{cucciati06}. Because of
the differences between the DEEP2 $R$-band and VVDS $I$-band selections, it
is possible to replicate only some of the VVDS subsamples. For $z < 0.9$,
we can construct volume--limited DEEP2 samples down to $M_B = -20$, while
for $z < 1.2$ we are able to construct a volume--limited sample matching
the VVDS $M_B < -21$ data set, and a very nearly volume--limited sample
with $M_B < -20.5$. Figures 7 and 8 of \citet{cucciati06} illustrate the
red fraction versus overdensity trends for the corresponding samples drawn
from VVDS.

\begin{figure*}[h!]
\centering
\includegraphics[scale=0.9]{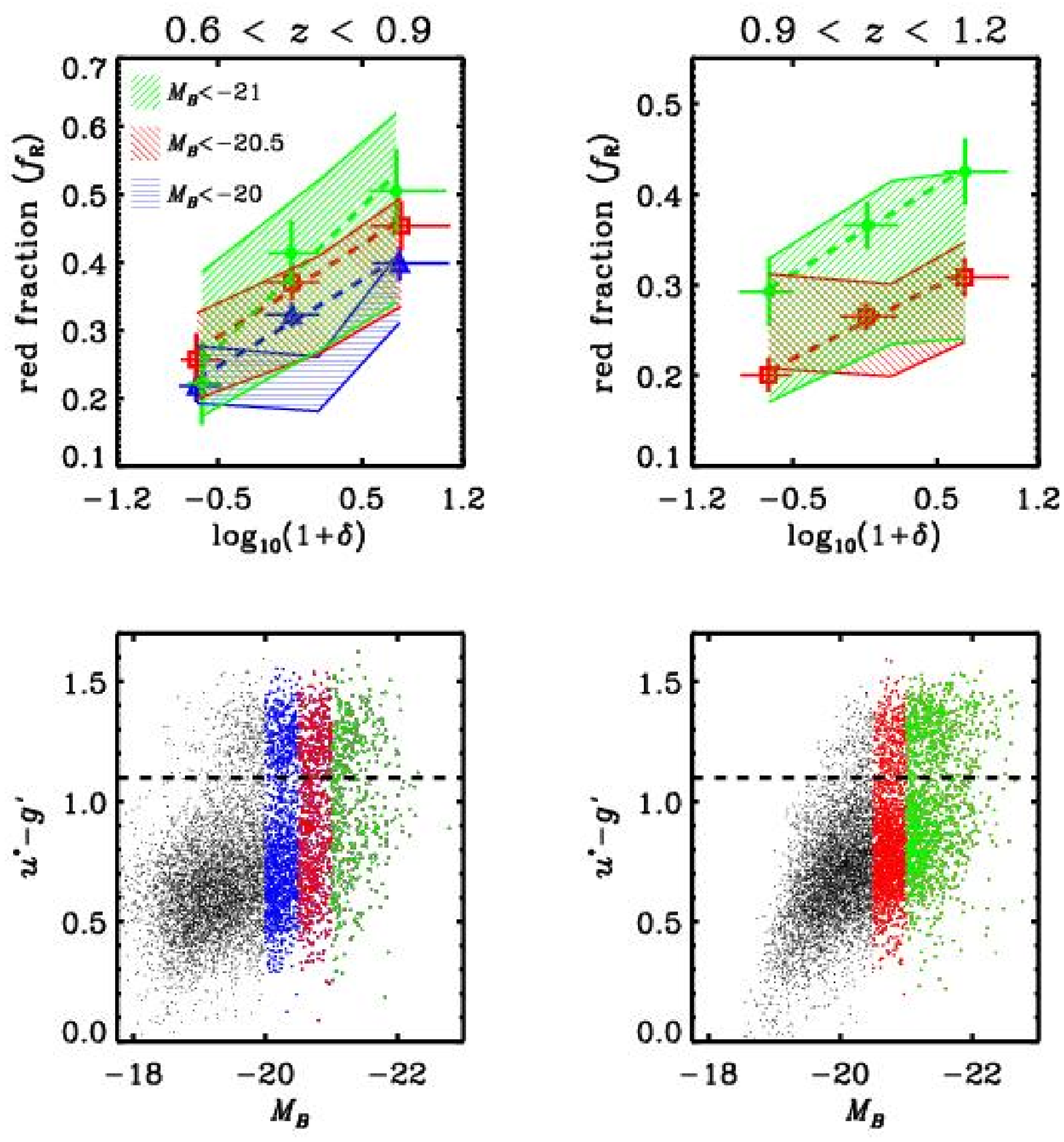}
\caption{As in Figure \ref{redfrac}, except here we plot the red fraction
  as a function of overdensity, $\log_{10}(1 + \delta_3)$, in two distinct
  redshift ranges, $0.6 < z < 0.9$ and $0.9 < z < 1.2$. The green circles,
  red squares, and blue triangles correspond to volume--limited samples
  with $M_B < -21$, $-20.5$, and $-20$, respectively. The shaded (or
  filled) regions in each panel show the 1--$\sigma$ range of the red
  fraction as a function of overdensity for the corresponding VVDS samples,
  as plotted in Figure 7 of \citet{cucciati06}; a key to the hatching
  patterns used is shown at upper left. 
  As discussed in the text, the 
  extreme--overdensity VVDS points have been mapped onto the corresponding
  extreme points from DEEP2, while the VVDS red fraction values remain
  unmodified. \emph{Bottom Row:} We plot the rest--frame color--magnitude
  relation, $u^{*}-g'$ versus $M_B$, for all DEEP2 objects in each redshift
  bin. The division between the red sequence and the blue cloud is given by
  the dashed line at $u^{*}-g' = 1.1$, following the definition of
  \citet{cucciati06}. The various magnitude--limited samples, $M_B < -21$,
  $-20.5$, and $-20$, are 
  denoted by the green, red, and blue colored symbols,
  respectively. Comparing like galaxy samples, the DEEP2 and VVDS results
  are generally in agreement, given the significantly larger errors for the
  \citet{cucciati06} data set. }
\label{ug_comp}
\end{figure*}

As shown in Figure \ref{ug_comp}, the DEEP2 results --- for volume--limited
samples at $0.6 < z < 0.9$ and $0.9 < z< 1.2$ --- are consistent with the
VVDS results for similarly--selected data sets. However, the errors on the
VVDS trends between red fraction and environment are significantly larger
($\gtrsim 2\times$ those for DEEP2). To make this figure, we map the data
points from Figure 7 of \citet{cucciati06} onto the DEEP2 results by
plotting the extreme-overdensity VVDS points at the same abcissa values as
the extreme-overdensity DEEP2 points. This is necessary because
citet{cucciati06} measures environments over much larger scales than we do
in DEEP2, and density contrasts on larger scales should be
smaller. However, locally, at least, the same trends are found using
environments measured on smaller and larger scales \citep{blanton06b}.

No scaling, however, is applied to the VVDS red--fraction values, as
plotted in Figure \ref{ug_comp}. The difference in normalization of $f_{\rm
  R}$ between DEEP2 and VVDS could be due to K--correction errors in either
sample, or, conceivably, due to a possible tendency of VVDS to fail to
obtain high--confidence redshifts for red galaxies as often as blue,
especially at $z \gtrsim 1$. 
From this comparison to the \citet{cucciati06} results, we conclude that
the color--density relation at $z > 0.9$, as observed by DEEP2, is
generally consistent with the VVDS measurements. However, due to their
larger uncertainties (likely principally due to their smaller sample size),
the color--density relation found here could not have been detected
significantly by \citet{cucciati06}.

Given that a perfect match between our samples and techniques and those of
\citet{cucciati06} is impossible,
especially due to the overdensity mapping applied in Figure \ref{ug_comp},
we also consider other possible reasons for why 
a trend present in DEEP2 could be missed within the VVDS data set. It is
quite possible that fundamental differences in the samples observed could
contribute to the differing conclusions at $z \sim 1$ from the DEEP2 and
VVDS analyses.
In particular, the VVDS sample includes a subset of objects to $I_{\rm AB}
= 24$, while DEEP2 samples down to a limit of $R_{\rm AB} = 24.1$. As a
consequence, while the surveys have comparable depths for the bluest
objects, VVDS is deeper for red galaxies at $z > 1$, which have $(R-I)_{AB}
\sim 1$--$1.5$. As a result, \citet{cucciati06} include fainter red
galaxies in their $f_{\rm R}$ measurements than we do in our standard
samples (in the volume--limited subsamples, the respective depths are
indentical, of course), and are able to study samples with
color--independent $M_B$ limits down to $-19.5$ (for $z < 1.2$).
As shown in Fig.\ \ref{ug_comp}, however, volume--limited DEEP2 samples
matching VVDS samples in luminosity range do exhibit a significant
color--density trend where the equivalent VVDS samples do not.

\begin{figure}[h!]
\centering
\plotone{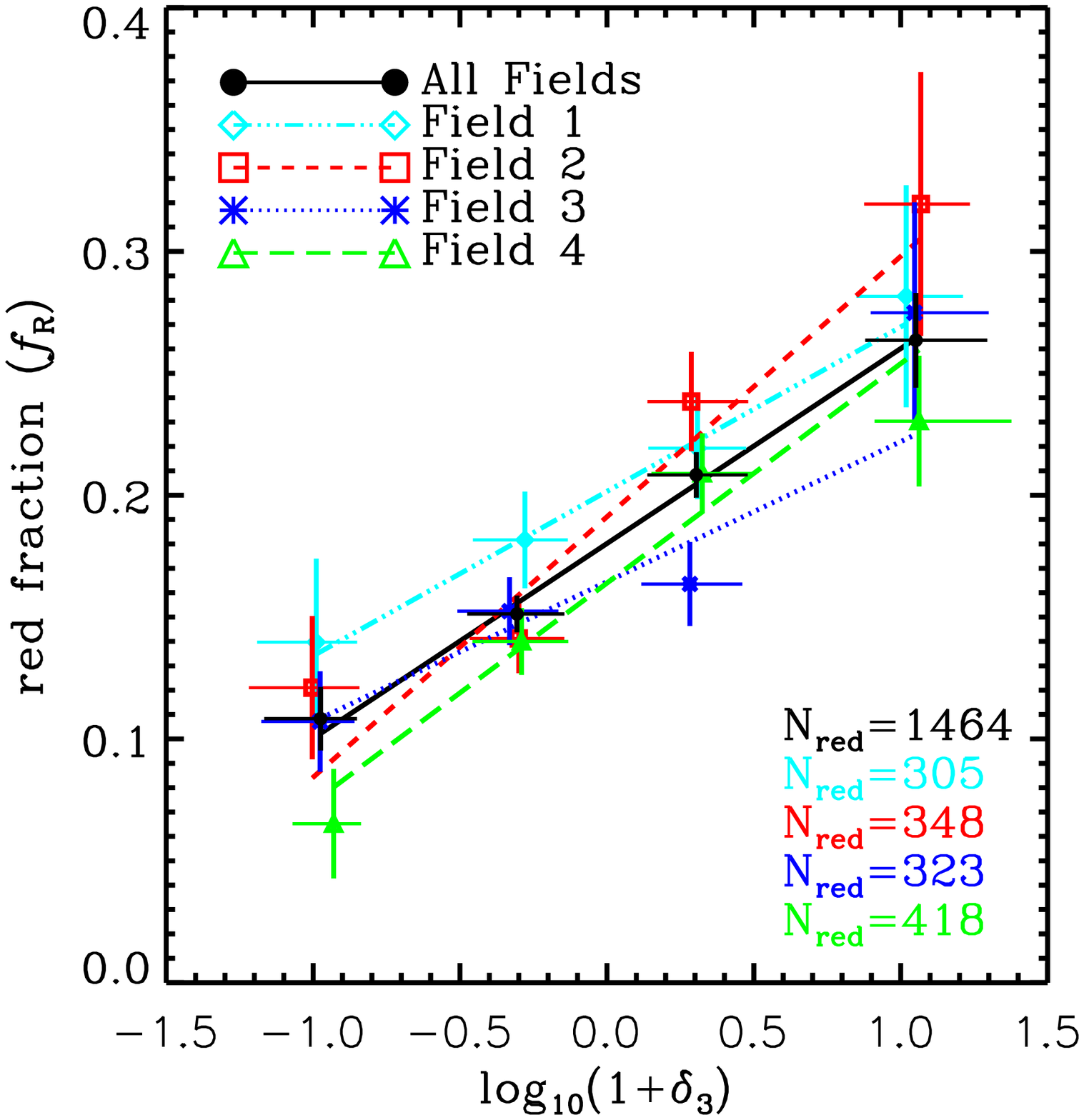}
\caption{Using the full galaxy sample (Sample A), we plot the red fraction
  as a function of overdensity, $\log_{10}{(1 + \delta_3)}$, at $0.8 < z <
  1$ in each of the four DEEP2 fields. The data points give the red
  fraction as a function of the median overdensity computed in four
  distinct bins of $\log_{10}{(1 + \delta_3)}$. The horizontal error bars
  run from the twenty--fifth percentile to the seventy--fifth percentile of
  the overdensity distribution in each bin. The vertical error bars give
  the 1--$\sigma$ uncertainty on the red fraction within each overdensity
  bin, given by Poisson statistics. The lines show linear--regression fits
  to the data points in each field. In the bottom right corner of the plot,
  we give the number of red galaxies within the $0.8 < z < 1$ redshift
  range in each field. The field--to--field variance in the $f_{\rm
    R}$--overdensity correlation is small, illustrating the small
  uncertainty due to cosmic variance. Cosmic variance mostly changes the
  overall red fraction in a given field and the relative abundance of the
  different environments, but not the strength of a trend.}
\label{cvplot}
\end{figure}

In addition to the \citet{cucciati06} sample being small (6,582 galaxies)
in comparison to our DEEP2 sample, yielding increased Poisson errors, it is
also limited to a single $0.7^{\circ} \times 0.7^{\circ}$ field, which
increases the uncertainties due to sample (or ``cosmic'') variance compared
to our sample. The differences between our results and those of
\citet{cucciati06}, however, are likely not attributable to cosmic variance
alone. As shown in Figure \ref{cvplot}, the variation in the $f_{\rm
  R}$--overdensity relation from field to field within the DEEP2 survey is
small, with each field yielding a color--density relation consistent with
that of the full sample. 
Cosmic variance mostly changes the overall red fraction in a given field
and the relative abundance of the different environments, but not the
strength of a trend.

In addition to smaller sample sizes, which yield larger errors in $f_{\rm
  R}$, there are other phenomena which could obscure a true correlation
between red fraction and overdensity in VVDS data preferentially at high
redshift. One possibility is that environment errors in the
\citet{cucciati06} sample are large compared to their environment bins at
$z > 0.9$. We note that the 5\%--95\% range of measured overdensities
increases with $z$ in their data set, as seen in Fig.\ 6 of
\citet{cucciati06}; this could reflect increasing errors in their
environment measures (in the DEEP2 data set, in contrast, the 5\%--95\%
range in measured overdensities is nearly independent of redshift).

Another difference is that \citet{cucciati06}
use a large smoothing
kernel ($\sim \! 5\ h^{-1}$ Mpc) in measuring environments. Estimating
overdensities on such large scales can be problematic due to the small
number of independent resolution elements across such a small field (at $z
\sim 1$, $0.7^{\circ}$ corresponds to $\sim 30\ h^{-1}$ comoving Mpc) and
due to the large percentage of the sample which falls within one smoothing
scale length of a survey boundary or edge ($> \! 50\%$ of the VVDS field by
area at $z \sim 1$). 
Differences between the true relationship connecting color and environment
on smaller scales and 5 Mpc scales \citep[as used by][]{cucciati06} could
also account for discrepancies between DEEP2 and VVDS results, but at $z
\sim 0$, at least, correlations between galaxy properties and environment
measurements on $\sim 1$ and $\sim 8$ Mpc scales appear to differ
significantly only in noise properties \citep{blanton06b}.

Finally, another effect that could dilute the measurements of
\citet{cucciati06} results from the fact that at $z \gtrsim 1$ --- the
redshift regime where a lack of a color--density relation is found in the
\citet{cucciati06} data --- the VVDS sample is dominated by lower--quality
VVDS redshifts \citep[${\rm flag} = 2$,][]{ilbert05}.
A number of tests have found that the ${\rm flag} = 2$ redshifts have a
nonnegligible error rate \citep[$\sim 20\%$, O.\ Ilbert, private
communication; C.\ Wolf, private communication;][]{lefevre05b}, which would
cause sources to be included in the \citet{cucciati06} sample at $z > 1$
incorrectly, with both erroneous K--corrections and erroneous overdensity
measurements. This contamination would dilute any true correlation between
rest--frame color and environment. In this paper, we use only
higher--confidence DEEP2 redshifts, which have an overall failure rate
below 2\% based upon tests with repeated observations \citep{davis06}.

While the color--density results of \citet{cucciati06} differ when compared
to the trends found in the full DEEP2 sample, a comparison of similar
samples (cf.\ Fig.\ \ref{ug_comp}) illustrates that the DEEP2 and VVDS
results are mostly in agreement where they overlap, given the significantly
larger errors for the \citet{cucciati06} data set, which are likely due to
the smaller VVDS sample size. The various sources of error discussed above
(cf.\ \S 4.1) could also be at play, though the dominant uncertainty
appears to be statistical in nature. Overall, both studies independently
find that the color--density relation grows weaker with increasing
redshift.

Although trends at $0.9 < z < 1.2$ are not significant for any of the VVDS
samples, at lower redshift \citet{cucciati06} conclude that the
color--density relation is stronger for brighter galaxies than faint ones.
In contrast, we do not observe significant variations in the color--density
trend for samples differing in $M_B$. That is, over the absolute--magnitude
ranges probed in Fig.\ \ref{ug_comp}, the slope of the $f_{\rm
  R}$--overdensity relation exhibits no statistically significant
dependence on luminosity in our measurements. The volume--limited
subsamples in Fig.\ \ref{ug_comp}, however, are overlapping rather than
independent in luminosity and only probe a limited range of $M_B$. Thus,
the lack of significant luminosity dependence to the color--density
relation within DEEP2 is not a strong statement. As shown by
\citet{cooper06a}, the dependence of mean environment on luminosity and
color is effectively separable at $z \sim 1$, which implies that the
dependence of the color--density relation on luminosity should not be
strong over the luminosity range covered by DEEP2.

Within the context of the larger picture of galaxy evolution, as discussed
in \S 4.2, a luminosity dependence to the evolution of the color--density
relation may well exist. Comparison of our results and those of clustering
studies at $z \sim 1$ \citep[e.g.,][]{coil04a, coil06} to clustering
measurements of bright galaxies at $z \sim 2$ \citep[e.g.,][]{quadri06}
lead to similar conclusions. As explained by \citet{cucciati06}, such a
result can be easily understood as an example of cosmic ``downsizing''
\citep{cowie96}, where the cessation of star formation occurs first in
high--luminosity galaxies or high--mass halos (i.e., high--density
environments). However, the methods used in this paper do not detect this
trend within the DEEP2 data set.

\subsection{Implications of the Observed Evolution in $f_{\rm R}(z)$}

A quite striking result from our work is the difference between the
evolution of the red fraction in under-- and overdense environments spanning
the redshift range $0.75 < z < 1.3$; while $f_{\rm R}$ in underdense
environments remains roughly constant with $z$ in our sample, we find that
the red fraction in dense environments decreases with $z$, such that it is
roughly equal to the fraction in underdense environments at $z \sim
1.3$. Studying the properties of group galaxies in DEEP2, \citet{gerke06b}
find similar evolution in the color--density relation from $0.7 \lesssim z
\lesssim 1.3$, with the fraction of blue galaxies in groups becoming
comparable to that of the field at $z \sim 1.3$. Given $\sim 1$ Gyr to move
from the blue cloud to the red sequence \citep[e.g.,][]{vandokkum01,
  balogh04b}, if the color--density relation first arises at $z \sim 1.35$,
then the physical mechanisms responsible must start to become operative at
$z \sim 1.7$ within a concordance cosmology.

This observed evolution in the color--density relation and the existence of
an epoch at which environment--dependent quenching initiates are both
consistent with the current theoretical picture in which the conversion of
blue galaxies into members of the red sequence occurs in dark matter halos
with mass greater than some critical value $(\sim {\rm a\ few} \times
10^{11}\ {\rm M}_{\sun})$. As halos pass this mass threshold in such
hot--flow/cold--flow accretion models, the infalling cold gas supply to the
central galactic disk is virial shocked and shut off such that the galaxy
will quickly burn its remaining fuel and redden \citep[e.g.,][]{birnboim03,
  keres05}. Within the models of \citet{croton06} and others \citep[see
also][]{bower06, cattaneo06, kang06}, low--energy AGN activity is included
to suppress the cooling of shocked gas and the recommencement of star
formation.

At yet higher redshifts $(z \gtrsim 2)$, even in halos above this critical
threshold, quenching does not occur since cooling is effectively able to
remove the pressure support behind the virial shock \citep{birnboim03}. At
later times, the evolution of the halo mass function \citep{jenkins01,
  springel05} in combination with the near redshift--independence of the
critical mass \citep{keres05, croton06, cattaneo06} leads to a continual
increase in the number of halos above the threshold mass. Given the
correlation between halo mass and environment in simulations
\citep[e.g.,][]{lemson99, maulbetsch06, wetzel06}, this thereby predicts an
evolution in the color--density relation (i.e., becoming stronger) at $z
\lesssim 1.3$. In this picture, the color--density or morphology--density
relation within cluster environments likely persists out to higher
redshifts, as cluster galaxies reside in the most massive halos, those
first to reach the critical mass at $z \sim 2$. As DEEP2 primarily samples
the more common, less massive groups, the color--density relation is still
weak at somewhat later times, $z \sim 1.3$.

In addition to predicting evolution of the color--density relation at $z <
1.3$, these hot--flow/cold--flow accretion models also provide an additional
mechanism by which to explain the presence of red galaxies residing in
underdense regions (cf.\ \S 3.2). While passive evolution or merging of
several smaller galaxies to form a fossil group are also viable physical
mechanisms by which to create such galaxies, the tail (at $\gtrsim {\rm a\
  few} \times 10^{11}\ {\rm M}_{\sun}$) of the halo mass function in
low--density environments naturally leads to the existence of at least some
true red--sequence galaxies in voids.

The measured evolution in the color--density relation as presented here
should be directly related to measurements of galaxy clustering at high
redshift. We predict that the clustering of $\sim L^{*}$ galaxies at $z
\gtrsim 1.3$ should depend only weakly on color. Studies of galaxy
clustering by spectral type or by color at high redshift
\citep[e.g.,][]{coil04a, coil04b, meneux06} have shown that blue,
star--forming galaxies at $z \sim 1$ are less strongly clustered than their
red counterparts. Such clustering work, however, spans a broad redshift
range (extending to redshifts less than unity) and the galaxy samples
utilized are dominated in number by galaxies at $z < 1.2$, where we still
find a significant color--density relation. Unfortunately, the DEEP2 sample
includes a modest number of galaxies $(\sim 5000)$ at $z > 1.2$, with very
few on the red sequence; we therefore lack the statistical power needed to
compute correlation strengths for subsamples by galaxy color at such
redshifts.

At yet higher redshifts $(z \gtrsim 2)$ there are indications that
clustering depends on color. Studies of UV--selected galaxies and red,
near--IR bright galaxies have found significant differences in clustering
strengths depending on the sample selection, with the latter being more
strongly clustered \citep[e.g.,][]{daddi03, adelberger05a,
  foucaud06}. Furthermore, within near--IR bright samples, the measured
correlation length depends significantly on apparent color (e.g., $R_{\rm
  AB} - K_{\rm Vega}$ or $J-K$), such that blue near--IR bright galaxies
cluster like UV--selected Lyman--break galaxies (LBGs), while the red near--IR
bright galaxies exhibit a correlation length larger by a factor of roughly
two \citep{adelberger05b, quadri06}.

The red, near--IR bright samples observed at early epochs, however, are
likely the antecedents of the rarest, most massive red galaxies today and
are not representative of the progenitors of DEEP2 galaxies; their
correlation length at $z \gtrsim 2$ already exceeds that of the brightest
and reddest DEEP2 galaxy samples. Presumably, if we were to study similarly
extreme samples at $z \sim 1.3$ \citep[e.g., galaxies in massive clusters
analogous to the candidates studied by][]{rosati04, mei06a, mei06b,
  bremer06}, we should find a significant color--density relation. Instead,
the progenitors of DEEP2 galaxies were bluer and/or fainter at $z \sim 2$
and resembled the LBGs more than those red galaxies which are most readily
observed at high redshift.

Thus, our observations of the evolution of the red fraction in low--density
and high--density environments do not contradict the current set of
clustering measurements at higher redshift $(z \sim 2)$ or observations of
massive clusters at $z > 1$. Like the cluster galaxies, the red, near--IR
selected, massive galaxies seen at high $z$, given their observed
clustering, are likely to reside in very massive halos, which should be the
first halos to cross the threshold quenching mass. In fact, estimates of
their star--formation rates indicate that they are actively forming stars in
large quantities \citep{daddi04}; hence, they do not appear to generally
have been quenched at $z \sim 2$, though more recent observations of
near--IR selected galaxies at $z \sim 2$ indicate relatively low
star--formation activity in at least a portion of the massive galaxy
population at high redshift \citep{labbe05, kriek06}.

\section{Summary \& Conclusions}

In this paper, we present a detailed study of the evolution in the
color--density relation at $0.4 < z < 1.35$. Using a sample of galaxies
drawn from the DEEP2 Galaxy Redshift Survey, we estimate the local
overdensity about each galaxy according to the projected $3^{\rm
  rd}$--nearest--neighbor surface density. From this, we measure the
evolution of the red fraction with environment across time. Our principal
results are as follows:

\begin{itemize}

\item We find that the color--density relation observed locally still exists
  at $z > 1$; the fraction of galaxies on the red sequence increases with
  local galaxy overdensity to nearly the redshift limits of the DEEP2
  survey.

\item At all epochs probed $(0.4 < z < 1.3)$, we find there exists a
  population of red, morphologically early--type galaxies residing in the
  the most underdense environments.

\item The color--density relation evolves with redshift, growing weaker with
  lookback time such that at $z \gtrsim 1.3$ there is no detectable
  dependence of galaxy color on local environment in the DEEP2 sample.

\item Our results support a picture in which the red sequence grew
  preferentially in dense environments (i.e., galaxy groups) at $z \lesssim
  1.5$. Clearly, the local environment plays an important role in
  ``nurturing'' galaxies, establishing the existence of correlations such
  as the morphology--density and color--density relation over cosmic
  time. The strength of evolutionary trends suggests that the correlations
  observed locally do not appear to have been imprinted (by ``nature'')
  upon the galaxy population during their epoch of formation.

\item Our findings imply that there should be little color dependence in
  the clustering of $\sim L^{*}$ galaxies at $z \gtrsim 1.3$.

\end{itemize}


\acknowledgments This work was supported in part by NSF grants AST--0071048
AST--0071198, AST--0507428, AST--0507483. J.A.N.\ and A.L.C.\ acknowledge
support by NASA through Hubble Fellowship grants HST--HF--01165.01--A and
HST--HF--01182.01--A, respectively, awarded by the Space Telescope Science
Institute, which is operated by AURA Inc.\ under NASA contract NAS
5--26555. S.M.F.\ would like to acknowledge the support of a Visiting
Miller Professorship at UC--Berkeley. M.C.C.\ would like to thank Greg
Wirth and all of the Keck Observatory staff for their help in the
acquisition of the Keck/DEIMOS data. In addition, the authors thank Olga
Cucciati, Chris Marinoni, and the VVDS team for helpful discussions related
to this work, for review of an early version of this text, and for
providing their data points for comparison. We also thank the referee,
Felipe Menanteau, for his insightful comments and suggestions for improving
this work.

We also wish to recognize and acknowledge the highly significant
cultural role and reverence that the summit of Mauna Kea has always
had within the indigenous Hawaiian community. It is a privilege to be
given the opportunity to conduct observations from this mountain.


\end{document}